\documentclass[12pt, a4paper]{article}
\usepackage[utf8]{inputenc}
\usepackage{dcolumn,lscape}%,setspace}
\usepackage{amsmath,longtable,multicol,dcolumn,tabularx,graphicx,amssymb}
\usepackage{exscale,amsthm,multirow,rotating,endfloat}
\usepackage{natbib}
\usepackage{bbm}
\usepackage{tikz,dsfont,float}
\usepackage[caption = false]{subfig}
\newcommand{\xvec}{\boldsymbol}
\newcommand{\xmat}{\mathbf}

\usetikzlibrary{patterns}
\begin{document}

\def\spacingset#1{\renewcommand{\baselinestretch}%
{#1}\small\normalsize} \spacingset{1}

%%%%%%%%%%%%%%%%%%%%%%%%%%%%%%%%%%%%%%%%%%%%%%%%%%%%%%%%%%%%%%%%%%%%%%%%%%%%%%

\title{\bf Estimation of the spatial weighting matrix for regular lattice data -- An adaptive lasso approach with cross-sectional resampling}
\author{Miryam S. Merk\\
\small{European University Viadrina, Frankfurt (Oder), Germany}\\
  Philipp Otto\\
    \small{Leibniz University Hannover, Germany}}
  \maketitle

\begin{abstract}
Spatial econometric research typically relies on the assumption that the spatial dependence structure is known in advance and is represented by a deterministic spatial weights matrix.
Contrary to classical approaches, we investigate the estimation of sparse spatial dependence structures for regular lattice data. In particular, an adaptive least absolute shrinkage and selection operator (lasso) is used to select and estimate the individual connections of the spatial weights matrix. To recover the spatial dependence structure, we propose cross-sectional resampling, assuming that the random process is exchangeable. The estimation procedure is based on a two-step approach to circumvent simultaneity issues that typically arise from endogenous spatial autoregressive dependencies. The two-step adaptive lasso approach with cross-sectional resampling is verified using Monte Carlo simulations. Eventually, we apply the procedure to model nitrogen dioxide ($\mathrm{NO_2}$) concentrations and show that estimating the spatial dependence structure contrary to using prespecified weights matrices improves the prediction accuracy considerably.
\end{abstract}

\noindent%
\vfill

% \newpage
\spacingset{1.45} % DON'T change the spacing!

\section{Introduction}
Modeling spatial dependencies in spatial or spatio-temporal data theoretically requires accounting for $n^2-n$ potential interactions between the $n$ spatial units of a sample, which raises two important challenges. First, the number of unknown connections grows quadratically with the sample size, which can become computationally demanding especially in the context of big data. Second, the identification suffers from an incidental parameter problem because the number of unknown parameters exceeds the sample size.

Classical econometric approaches typically replace unknown spatial dependencies with the linear combination of a deterministic spatial weighting matrix, which must be prespecified, and an unknown scalar parameter reflecting the strength of the spatial dependence. Of course, the results of such models are only valid if the weights matrix has been correctly specified. However, in most applications the true underlying dependence structure is unknown. Only in very few exceptional cases, could one assume that the spatial dependence structure is
known (e.g., if the underlying physical process is well-understood). Thus, the spatial weights matrix is often based on distance measures between the individual locations, which involves geographic proximity (cf. \citealt{Cliff73, Anselin88a, Ertur07}) or other types of adjacency, such as social (see, e.g., \citealt{Cohen99}) or economic (see, e.g., \citealt{Bodson75, Besner02}) characteristics.
However, this approach has raised considerable criticism, especially because different weighting matrices can lead to very different and, therefore, contradicting estimation results (cf. \citealt{Mizruchi08, Smith09}). In particular, the results must be interpreted conditioned on the assumed weighting scheme (cf. \citealt{Debarsy12}).

Consequently, a variety of alternative approaches have been suggested that involve parametric or semi-parametric estimation procedures (see, e.g., \citealt{Pinkse02}) or matrix selection from a set of candidate schemes based on goodness-of-fit criteria, such as maximized log-likelihood values (cf. \citealt{Stakhovych09}). However, selection procedures that are based on competing matrix specifications require that the true weights matrix is among the candidate schemes. Alternatively, \cite{Bhattacharjee13} proposed estimating the weights from an estimated spatial autocovariance matrix. However, this approach uses panel data under the constraint of symmetry and a finite sample size of $n$. Hence, the number of unknown spatial connections or weights is reduced by half, but irregular patterns, such as anisotropy or time-varying spatial dependencies, are not covered. In addition to symmetry, other identifying assumptions have been proposed to estimate the individual spatial weights. For instance, the least shrinkage and selection operator (lasso) has been used by \cite{Tibshirani96} to estimate sparse spatial weighting matrices. \cite{Ahrens15} suggested identifying an approximately sparse spatial weights matrix using a two-stage lasso estimation procedure.
\cite{Lam19} proposed estimating the spatial weights matrix from a linear combination of different specifications adjusted using a potentially sparse matrix. Moreover, \cite{Otto18} discussed an adaptive lasso procedure to estimate the spatial dependence and an unknown number of possible structural breaks simultaneously.

However, all these approaches have in common that they require spatio-temporal data where the spatial dependence is not varying over time. To the best of our knowledge, little or no attention has been paid to purely spatial models where the number of spatial links exceeds the sample size and where no dependence structure must be assumed in advance. \cite{Zhu10} proposed penalized maximum likelihood (ML) estimators to select covariates and a neighborhood structure for lattice data in the context of conditional and simultaneous spatial error models. More precisely, the unknown spatial weights matrix is represented by a linear combination of individual matrices reflecting the neighborhood sets of different orders.

Contrary to these approaches, we suggest estimating all relevant spatial connections
for purely spatial regular lattice data individually under the identifying
assumption of sparsity and exchangeability.
More precisely, the spatial weights structure and regressors are selected and estimated by the adaptive lasso approach proposed by \cite{Zou06} as an extension of the original lasso that enables individual shrinking. In addition, cross-sectional resampling is applied to estimate the relevant spatial connections assuming that the weighting scheme is not varying over space. Thus, instead of estimating the spatial dependence structure over an entire (possibly high-dimensional) dataset, smaller exchangeable subsamples are considered to investigate spatial spillover effects. To avoid endogeneity issues arising due to the dependent variable simultaneously serving as an explanatory variable, we propose a two-step approach incorporating instrumental variables in the first step.

The rest of this paper is organized as follows. Section \ref{sec:theory} describes the theoretical framework, including the spatial autoregressive regression model, regularity assumptions, and the two-step adaptive lasso estimation procedure. The results obtained from the Monte Carlo simulations on the performance of the spatial weights estimates are reported in Section \ref{sec:MC}. Furthermore, we apply the two-step adaptive lasso approach to model nitrogen dioxide ($\mathrm{NO_2}$) concentrations. Finally, the last section presents the conclusion.

\section{Theoretical Model}\label{sec:theory}
Let $\{Y(\xvec{s}): \xvec{s} \in D\}$ be a spatial process at known locations $\xvec{s}$ in the set $D$, which is a subset of the $d$-dimensional real numbers $\mathbb{R}^d$. Moreover, $\xvec{s}$ may vary discretely or continuously over $D$. Spatial data can be generally classified into three categories, namely, spatial point patterns, geostatistical (or continuous) processes, and lattice data. For the latter case, the process is observed on regular or irregular grids in the two-dimensional space. 

In this paper, the focus is on spatial autoregressive models. For such processes, the observations in all regions $\{\xvec{s}_1, \dots, \xvec{s}_n\}$ denoted by the vector $\xvec{Y} = Y(\xvec{s}_i)_{i=1,\dots,n}$ are spatially interdependent. These dependencies are commonly characterized by an $n \times n$ spatial weights matrix $\xmat{W} = (w_{ij})_{i,j = 1,\dots,n}$, which relates each spatial unit to all other locations. More precisely, the $i$-th row illustrates how the corresponding observation on the dependent variable $Y(\xvec{s}_i)$ is influenced by observations in all other regions. Thus, the model equation of a spatially autoregressive response variable can be specified as follows:
\begin{equation}
\xvec{Y} = \xmat{W} \xvec{Y} + \xmat{X} \xvec{\beta} + \xvec{\epsilon},
\end{equation}
or as the reduced form:
\begin{equation}
\xvec{Y} = (\xmat{I} - \xmat{W})^{-1} (\xmat{X} \xvec{\beta} + \xvec{\epsilon}),
\end{equation}
where $\xmat{X}$ is an $n \times k$ matrix comprising $k$ explanatory variables, $\xmat{I}$ is the identity matrix of dimension $n$, and $\xvec{\epsilon}$ is an $n$-dimensional vector of independently and identically distributed residuals with zero mean and positive finite variance $\sigma^2$. The matrix $(\xmat{I} - \xmat{W})^{-1}$ represents a spatial multiplier effect that transfers changes in response variables to other areas that may not even be connected through $\xmat{W}$ (cf. \citealt{LeSage2009}). In particular, representing the matrix as a series expansion (i.e., $(\xmat{I} - \xmat{W})^{-1} = \xmat{I} + \xmat{W} + \xmat{W}^2 + \xmat{W}^3 + \dots$) elucidates the spatial spillovers from first, second, and higher-order neighbors.

The spatial weights matrix is a nonnegative matrix reflecting the spatial dependence structure. The diagonal elements of $\xmat{W}$ are zero (i.e., $w_{ii} = 0$ for all $i$) because no location is a neighbor of itself. Moreover, matrix $(\xmat{I} - \xmat{W})$ must be non-singular, and the row and column sums of $\xmat{W}$ and $(\xmat{I} - \xmat{W})^{-1}$ are bounded in absolute value (cf. \citealt{Kelejian1999}). More formally, $\left\lVert \xmat{W} \right\rVert_1 < a$ and $\left\lVert \xmat{W} \right\rVert_\infty < a$, and $a/n \to 0$ as $n \to \infty$. These assumptions restrict the cross-sectional correlation to a controllable degree and ensure that the spatial dependence decreases with increasing distance. This also involves spatial spillover effects from higher-order neighbors that diminish with increasing orders.

Finally, the identifying assumption of the sparsity of the spatial weights matrix is employed. Thus, we assume that each location is influenced by only a few other locations lying within a certain distance, which is described in more detail below.

\subsection{Estimation procedure}
Classical least squares estimators are biased in the context of endogenous spatial dependencies. Consistent procedures, such as the ML approach (cf. \citealt{Ord75, Anselin88a, Lee04}) or generalized method of moments (cf. \citealt{Kelejian1999}), circumvent simultaneity issues arising due to interdependence among neighboring locations. Alternatively, \cite{Kelejian98a} proposed a two-stage least squares estimation procedure, which is computationally less demanding than the ML approach and does not require prior distributional assumptions. In addition, \cite{Ahrens15} adapted a two-step estimation procedure for spatio-temporal data to estimate spatial weights, where the number of instruments can be larger than the sample size, assuming approximate sparsity.

\begin{figure}
\begin{center}
\begin{tikzpicture}[scale=0.50]%[x=20pt,y=10pt,z=5pt]
%%%%%%%%%%%%%%%%%% 2 dimensional %%%%%%%%%%%%%%%%%%%%
\foreach \a / \b / \c / \d / \opac / \col in {  -22 / -10  / 18 / 2  / 0   / 80, % locations valid for sample (opacity = 0!!)
                                                -24 / -22  / 0  / 20 / 0.8 / 80, % locations excluded for sample
                                                -4  / -6  / 0  / 20 / 0.8 / 80,
                                                -22 / -6  / 18 / 20 / 0.8 / 80,
                                                -22 / -6  / 0  / 2  / 0.8 / 80}
                                                {
    \coordinate (A11) at (\a,\d,5);
    \coordinate (A12) at (\b,\d,5);
    \coordinate (A13) at (\a,\c,5);
    \coordinate (A14) at (\b,\c,5);
\draw[fill=black!\col,opacity=\opac, pattern=north east lines]  (A11) -- (A12) -- (A14) -- (A13); %
}

%%%%%%%%%%%%%%%%% colors %%%%%%%%%%%%%%%%%%%%%%%%

\draw[fill = orange, opacity = 0.8] (-22,14,5) -- (-20,14,5) -- (-20,12,5) -- (-22,12,5);
\draw[fill = orange, opacity = 0.2] (-24,16,5) -- (-18,16,5) -- (-18,10 ,5) -- (-24,10 ,5);
\draw[fill = blue, opacity = 0.8]   (-12,8 ,5) -- (-10,8 ,5) -- (-10,6 ,5) -- (-12,6 ,5);
\draw[fill = blue, opacity = 0.2]   (-14,10,5) -- (-8 ,10,5) -- (-8 ,4 ,5) -- (-14,4 ,5);
%\draw[fill = blue, opacity = 0.5] (-8,6,5) -- (-6,6,5) -- (-6,4,5) -- (-8,4,5);
%\draw[fill = blue, opacity = 0.5] (-8,8,5) -- (-6,8,5) -- (-6,6,5) -- (-8,6,5);

%%%%%%%%%%%%%%%%% true dependence %%%%%%%%%%%%%%%%

\draw[<-, color=orange!60!black!90, solid, line width=1pt] (-21,13,5) -- (-21,15,5);
\draw[<-, color=orange!60!black!90, solid, line width=1pt] (-21,13,5) -- (-19,13,5);
\draw[<-, color=orange!60!black!90, solid, line width=1pt] (-21,13,5) -- (-19,15,5);

\draw[<-, color=blue!60!black!90, solid, line width=1pt] (-11,7,5) -- (-11,9,5);
\draw[<-, color=blue!60!black!90, solid, line width=1pt] (-11,7,5) -- (-9,7,5);
\draw[<-, color=blue!60!black!90, solid, line width=1pt] (-11,7,5) -- (-9,9,5);

\draw[color=blue!60!black!90, solid, line width=2pt]  (-12,10,5) -- (-8,10,5) -- (-8,6,5) -- (-10,6,5) -- (-10,8,5) -- (-12,8,5) -- (-12,10,5);

\draw[color=orange!60!black!90, solid, line width=2pt]  (-22,16,5) -- (-18,16,5) -- (-18,12,5) -- (-20,12,5) -- (-20,14,5) -- (-22,14,5) -- (-22,16,5);

%%%%%%%%%%%%%%%%% grid %%%%%%%%%%%%%%%%%%%%%%%%
\foreach \x in {-4,-6,-8,-10,-12,-14,-16,-18,-20,-22,-24} % Vorderseite
    \foreach \y in {0,2,4,6,8,10,12,14,16,18,20}
\draw[color=black, solid, line width=0.5pt]  (\x,0,5) -- (\x,20,5) (-4,\y,5) -- (-24,\y,5);

% \draw[color=black, solid, line width=2pt]  (-22,0,5) -- (-6,20,5) (-4,18,5) -- (-24,0,5);

%%%%%%%%%%%%%%%%% axis %%%%%%%%%%%%%%%%%%%%%%%%
\draw[<->] (-26,-1,5) -- (-26,21,5); % node[above] {$y$}
\draw[<->] (-25,-2,5) -- (-3,-2,5); % node[right] {$x$}
\foreach \x / \lab in {-5/10,-7/9,-9/8,-11/7,-13/6,-15/5,-17/4,-19/3,-21/2,-23/1}{
    \draw[shift={(\x,-2,5)},color=black] (0pt,0pt) -- (0pt,-3pt) node[below] {$\lab$};
}
\foreach \y / \lab in {1/1,3/2,5/3,7/4,9/5,11/6,13/7,15/8,17/9,19/10}{
    \draw[shift={(-26,\y,5)},color=black] (0pt,0pt) -- (-3pt,0pt) node[left] {$\lab$};
}
%%%%%%%%%%%%%%%%%% legend %%%%%%%%%%%%%%%%%%%%%%%
\draw (-24,-5,5) node [circle, fill=black!100, draw=black, pattern=north east lines]{};
\draw (-15,-5,5) node {Locations that are not used for resampling};

\draw (-24,-6,5) node [circle, fill=orange, draw=black]{};
\draw (-22.5,-6,5) node {$\xvec{s}_{*_1}$};
\draw (-20,-6,5) node [circle, fill=orange, draw=black, opacity = 0.2]{};
\draw (-15,-6,5) node {$N_{*_1} = (\xvec{s}_{*_1}^{(-1)}, \ldots, \xvec{s}_{*_1}^{(8)})$};
\draw (-9,-6,5) node [circle, draw=orange!60!black!90, solid, line width=2pt]{};
\draw (-5.5,-6,5) node {True links $N^*_{*_1}$};
\draw (-24,-7,5) node [circle, fill=blue, draw=black]{};
\draw (-22.5,-7,5) node {$\xvec{s}_{*_2}$};
\draw (-20,-7,5) node [circle, fill=blue, draw=black, opacity = 0.2]{};
\draw (-15,-7,5) node {$N_{*_2} = (\xvec{s}_{*_2}^{(-1)}, \ldots, \xvec{s}_{*_2}^{(8)})$};
\draw (-9,-7,5) node [circle, draw=blue!60!black!90, solid, line width=2pt]{};
\draw (-5.5,-7,5) node {True links $N^*_{*_1}$};
\end{tikzpicture}
\end{center}
\caption[Cross-sectional resampling (first step)]{Cross-sectional resampling (first step). Two sample locations $\xvec{s}_{*_1}$ and $\xvec{s}_{*_2}$ are highlighted in orange and blue, respectively, including their eight nearest locations representing the sets $N_i$. The true links, $N_i^*$, are drawn by arrows. The setting has an anisotropic dependence from the north-east.}\label{fig:setting1}
\end{figure}
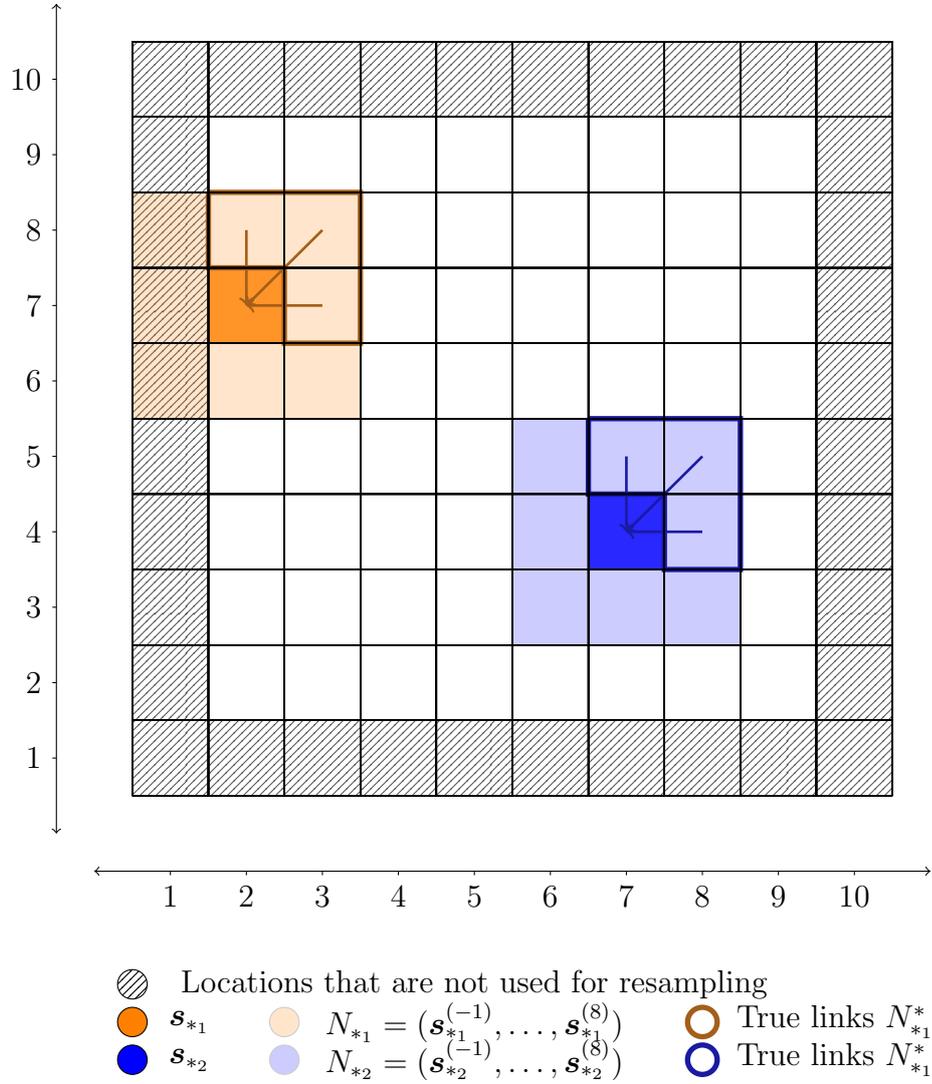

In contrast to spatio-temporal data, where the spatial dependence structure may be recovered by repeated observations over time, we propose using cross-sectional resampling for purely spatial processes. Suppose that all true connections of the $i$-th location are situated within a smaller subset $N_i$ of the $m$ nearest locations of $\xvec{s}_i$. Let $N_i^* \subset N_i$ be the set of the true links with $|N_i^*| = q$. Moreover, suppose that the number of positive weights $q$ is much smaller than $m$, such that the resulting dependence structure is sparse.  Then, we randomly select $r$ spatial locations for which the set of the $m$ nearest locations is complete. This excludes border units that may be included in the set of potential neighbors but are not sampled as dependent variables. Eventually, $\tilde{\xvec{Y}}$ denotes the vector of sampled observations $(Y(\xvec{s}_{*_1}), \ldots, Y(\xvec{s}_{*_r}))'$.
Figure \ref{fig:setting1} illustrates the cross-sectional resampling approach using two randomly selected locations with $m = 8$ nearest locations and $q = 3$ true neighbors to the north, north-east, and east.

To identify the spatial dependence structure, we assume that the random process is exchangeable (i.e., the joint distribution of the random variables in $N_i$ and of their spatial interrelations is the same for all $i$). Accordingly, a unique parameter vector $\xvec{w} = (w_1, \dots, w_m)'$ exists, reflecting how each location is affected by its $m$ nearest locations. In other words, $\xvec{w}$ represents one row of $\xmat{W}$, excluding the diagonal entry and all entries outside $N_i$, which are assumed to be zero. The full spatial weights matrix can be reconstructed from $\xvec{w}$, assuming that the individual locations of the spatial observations are known. This assumption is in line with standard econometric specifications presuming the existence of a universal spatial weighting scheme that applies to all cross-sectional locations, like the $q$ nearest neighbors, binary contiguity, and inverse distance weighting.

\subsection{Two-step lasso estimator}
In general, our estimation procedure is a two-step adaptive lasso approach that allows recovering the spatial dependence structure based on cross-sectional resampling. The first step involves finding suitable instruments to predict the endogenous variables $\tilde{\xvec{Y}}$. In the second step, the response variables are replaced with their predicted values to circumvent simultaneity issues, and all parameters of the full model are estimated.

\subsubsection{First step: IV regression}
In the first step, we estimate $\tilde{\xvec{Y}}$ using $l$ instruments:
\begin{equation}\label{firststep}
\tilde{\xvec{Y}} = \tilde{\xmat{Z}} \xvec{\theta}  + \xvec{\nu},
\end{equation}
where $\tilde{\xmat{Z}}$ is the corresponding $r \times l$-dimensional matrix of instruments at the $r$ sampled locations, and $\xvec{\theta}$ is the corresponding $l$-dimensional vector of coefficients with $l \geq 1$. It is assumed that $E[\epsilon(\xvec{s}_{*_i})|\xvec{Z}_p] = 0$ for all $i=1,\dots,r$ and $p=1,\dots,l$ (i.e., the instruments are independent of the regression residuals but correlated with the response variable or endogenous variables in general).
In contrast to the proposal by \cite{Kelejian98a} that the instruments should be composed of exogenous regressors $\xmat{X}$ and first and higher-order spatial lags $(\xmat{W} \xmat{X}, \xmat{W}^2\xmat{X}, \dots)$ using prespecified spatial weights, the spatial weights matrix is unknown in our case. Thus, as instruments, we take all exogenous regressors at location $\xvec{s}_i$ and its $m$ nearest locations. Consequently, there are $l = k + mk$ instruments with $m \ll n$. For simplicity, we assume that the size of the subset of potential neighbors $|N_i| = m$ is the same in the first and second steps.

The first-step adaptive lasso estimator solves the following:
\begin{equation}
\hat{\xvec{\theta}} = \arg\min_{\xvec{\theta}} \left\lVert \tilde{\xvec{Y}} - \tilde{\xmat{Z}} \xvec{\theta} \right\rVert^2_2 + \lambda_1 \left\lVert \xvec{\psi}_1 \circ \xvec{\theta} \right\rVert_1 \, .
\end{equation}
The objective function consists of two parts, where the first term minimizes the residual sum of squares between observations on the response variables and corresponding instruments.
Using the estimated coefficients $\hat{\xvec{\theta}}$ and the $n \times l$ matrix $\xmat{Z}$, we can predict the vector $\xvec{Y}$, except for the locations at the edges of the random field. Eventually, these predicted values denoted by $\breve{Y}(\xvec{s}_i)$ are used in the second step to avoid endogeneity.

\subsubsection{Second step: Full model}
\begin{figure}
\begin{center}
\begin{tikzpicture}[scale=0.50]%[x=20pt,y=10pt,z=5pt]
%%%%%%%%%%%%%%%%%% 2 dimensional %%%%%%%%%%%%%%%%%%%%
\foreach \a / \b / \c / \d / \opac / \col in {  -22 / -6  / 18 / 2  / 0   / 80, % locations valid for sample (opacity = 0!!)
                                                -24 / -20  / 0  / 20 / 0.8 / 80, % locations excluded for sample
                                                -4  / -8  / 0  / 20 / 0.8 / 80,
                                                -20 / -8  / 16 / 20 / 0.8 / 80,
                                                -20 / -8  / 0  / 4  / 0.8 / 80}
                                                {
    \coordinate (A11) at (\a,\d,5);
    \coordinate (A12) at (\b,\d,5);
    \coordinate (A13) at (\a,\c,5);
    \coordinate (A14) at (\b,\c,5);
\draw[fill=black!\col,opacity=\opac, pattern=north east lines]  (A11) -- (A12) -- (A14) -- (A13); %
}

    \coordinate (A11) at (-22,2,5);
    \coordinate (A12) at (-6,2,5);
    \coordinate (A13) at (-22,18,5);
    \coordinate (A14) at (-6,18,5);
\draw[color=black!40, solid, line width=2pt]  (A11) -- (A12) -- (A14) -- (A13) -- (A11);

%%%%%%%%%%%%%%%%% colors %%%%%%%%%%%%%%%%%%%%%%%%

\draw[fill = orange, opacity = 0.8] (-14,12,5) -- (-12,12,5) -- (-12,10,5) -- (-14,10,5);
\draw[fill = orange, opacity = 0.2] (-16,14,5) -- (-10,14,5) -- (-10,8 ,5) -- (-16,8 ,5);
\draw[fill = blue, opacity = 0.8]   (-10,8 ,5) -- (-8,8 ,5) -- (-8,6 ,5) -- (-10,6 ,5);
\draw[fill = blue, opacity = 0.2]   (-12,10,5) -- (-6 ,10,5) -- (-6 ,4 ,5) -- (-12,4 ,5);
%\draw[fill = blue, opacity = 0.5] (-8,6,5) -- (-6,6,5) -- (-6,4,5) -- (-8,4,5);
%\draw[fill = blue, opacity = 0.5] (-8,8,5) -- (-6,8,5) -- (-6,6,5) -- (-8,6,5);

\draw (-13,11,5) node {\tiny{$Y(\xvec{s}_{*_1})$}};
\draw (-15,13,5) node {\tiny{$\breve{Y}(\xvec{s}_{*_1}^{(1)})$}};
\draw (-13,13,5) node {\tiny{$\breve{Y}(\xvec{s}_{*_1}^{(2)})$}};
\draw (-11,13,5) node {\tiny{$\breve{Y}(\xvec{s}_{*_1}^{(3)})$}};
\draw (-15,11,5) node {\tiny{$\breve{Y}(\xvec{s}_{*_1}^{(4)})$}};
\draw (-11,11,5) node {\tiny{$\breve{Y}(\xvec{s}_{*_1}^{(5)})$}};
\draw (-15,9,5) node {\tiny{$\breve{Y}(\xvec{s}_{*_1}^{(6)})$}};
\draw (-13,9,5) node {\tiny{$\breve{Y}(\xvec{s}_{*_1}^{(7)})$}};
\draw (-11,8.5,5) node {\tiny{$\breve{Y}(\xvec{s}_{*_1}^{(8)})$}};

\draw (-9,7,5) node {\tiny{$Y(\xvec{s}_{*_2})$}};
\draw (-11,9.5,5) node {\tiny{$\breve{Y}(\xvec{s}_{*_2}^{(1)})$}};
\draw (-9,9,5) node {\tiny{$\breve{Y}(\xvec{s}_{*_2}^{(2)})$}};
\draw (-7,9,5) node {\tiny{$\breve{Y}(\xvec{s}_{*_2}^{(3)})$}};
\draw (-11,7,5) node {\tiny{$\breve{Y}(\xvec{s}_{*_2}^{(4)})$}};
\draw (-7,7,5) node {\tiny{$\breve{Y}(\xvec{s}_{*_2}^{(5)})$}};
\draw (-11,5,5) node {\tiny{$\breve{Y}(\xvec{s}_{*_2}^{(6)})$}};
\draw (-9,5,5) node {\tiny{$\breve{Y}(\xvec{s}_{*_2}^{(7)})$}};
\draw (-7,5,5) node {\tiny{$\breve{Y}(\xvec{s}_{*_2}^{(8)})$}};

%%%%%%%%%%%%%%%%% grid %%%%%%%%%%%%%%%%%%%%%%%%
\foreach \x in {-4,-6,-8,-10,-12,-14,-16,-18,-20,-22,-24} % Vorderseite
    \foreach \y in {0,2,4,6,8,10,12,14,16,18,20}
\draw[color=black, solid, line width=0.5pt]  (\x,0,5) -- (\x,20,5) (-4,\y,5) -- (-24,\y,5);

% \draw[color=black, solid, line width=2pt]  (-22,0,5) -- (-6,20,5) (-4,18,5) -- (-24,0,5);

%%%%%%%%%%%%%%%%% axis %%%%%%%%%%%%%%%%%%%%%%%%
\draw[<->] (-26,-1,5) -- (-26,21,5); % node[above] {$y$}
\draw[<->] (-25,-2,5) -- (-3,-2,5); % node[right] {$x$}
\foreach \x / \lab in {-5/10,-7/9,-9/8,-11/7,-13/6,-15/5,-17/4,-19/3,-21/2,-23/1}{
    \draw[shift={(\x,-2,5)},color=black] (0pt,0pt) -- (0pt,-3pt) node[below] {$\lab$};
}
\foreach \y / \lab in {1/1,3/2,5/3,7/4,9/5,11/6,13/7,15/8,17/9,19/10}{
    \draw[shift={(-26,\y,5)},color=black] (0pt,0pt) -- (-3pt,0pt) node[left] {$\lab$};
}
%%%%%%%%%%%%%%%%%% legend %%%%%%%%%%%%%%%%%%%%%%%

\draw (-24,-5,5) node [circle, fill=black!100, draw=black, pattern=north east lines]{};
\draw (-11.5,-5,5) node {Locations that are not used for resampling in the second step};

\draw (-24,-6,5) node [circle, draw=black!40, line width=2pt]{};
\draw (-24,-6,5) node [circle, draw=black, line width=0.5pt]{};
\draw (-12,-6,5) node {Locations for which $\breve{Y}(\xvec{s})$ can be predicted in the first step};

\draw (-24,-7,5) node [circle, fill=orange, draw=black]{};
\draw (-22.5,-7,5) node {$\xvec{s}_{*_1}$};
\draw (-20,-7,5) node [circle, fill=orange, draw=black, opacity = 0.2]{};
\draw (-15,-7,5) node {$N_{*_1} = (\xvec{s}_{*_1}^{(-1)}, \ldots, \xvec{s}_{*_1}^{(8)})$};
\draw (-24,-8,5) node [circle, fill=blue, draw=black]{};
\draw (-22.5,-8,5) node {$\xvec{s}_{*_2}$};
\draw (-20,-8,5) node [circle, fill=blue, draw=black, opacity = 0.2]{};
\draw (-15,-8,5) node {$N_{*_2} = (\xvec{s}_{*_2}^{(-1)}, \ldots, \xvec{s}_{*_2}^{(8)})$};
\end{tikzpicture}
\end{center}
\caption[Cross-sectional resampling (second step)]{Cross-sectional resampling (second step). Two sample locations $\xvec{s}_{*_1}$ and $\xvec{s}_{*_2}$ are highlighted in orange and blue, respectively, including their eight nearest locations representing the sets $N_i$. The locations for which $\breve{Y}(\xvec{s})$ could be predicted by the instrumental variables in the first step coincide with all locations that are used for resampling in the first step (Figure \ref{fig:setting1}). }\label{fig:setting2}
\end{figure}
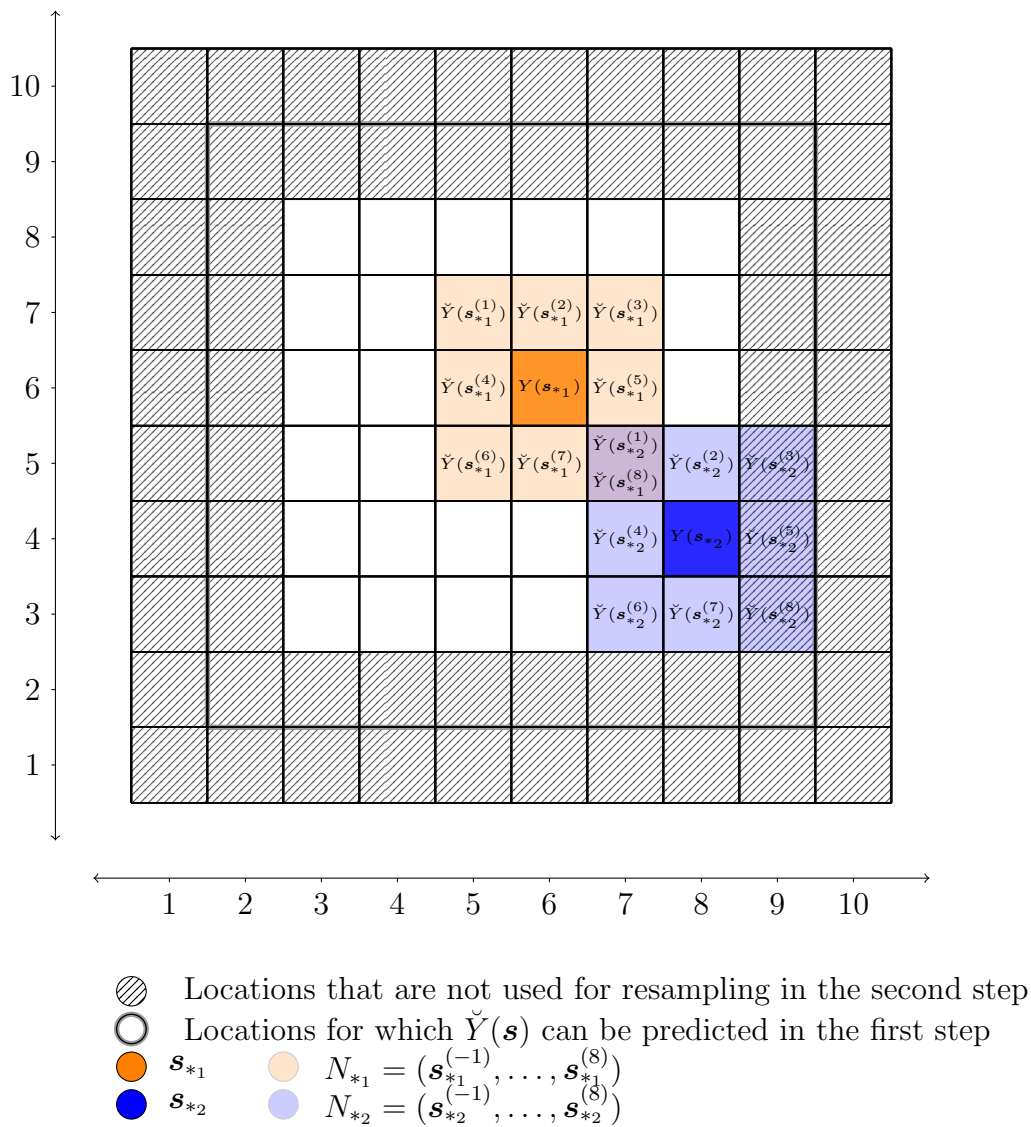

Let $\breve{\xmat{Y}}$ denote the $r \times m$ matrix of predicted values obtained from the first-step estimation:
\begin{equation}
\breve{\xmat{Y}} =
\left(
\begin{array}{ccc}
\breve{Y}(\xvec{s}_{*_1}^{(1)})  & \ldots & \breve{Y}(\xvec{s}_{*_1}^{(m)}) \\
\vdots & \ddots  & \vdots \\
\breve{Y}(\xvec{s}_{*_r}^{(1)}) & \ldots & \breve{Y}(\xvec{s}_{*_r}^{(m)}) \\
\end{array}
\right)
\end{equation}
with $\breve{Y}(\xvec{s}_{*_1}^{(1)}), \ldots, \breve{Y}(\xvec{s}_{*_1}^{(m)})$ as the $m$-nearest neighbors of the first sampled location $\xvec{s}_{*_1}$.
Then, the second-step adaptive lasso estimator solves the following:
\begin{equation}
(\hat{\xvec{\beta}}, \hat{\xvec{w}})' = \arg\min_{(\xvec{\beta}, \xvec{w})'} \left\lVert \tilde{\xvec{Y}} - \tilde{\xmat{X}} \xvec{\beta} - \breve{\xmat{Y}} \xvec{w} \right\rVert^2_2 + \lambda_2 (\left\lVert \xvec{\psi}_{2,\beta} \circ \xvec{\beta} \right\rVert_1 + \left\lVert \xvec{\psi}_{2,w} \circ \xvec{w} \right\rVert_1)
\end{equation}
\begin{equation*}
\text{s.t.} \xvec{w} \geq 0 \text{ and } \left\lVert \xvec{w} \right\rVert_1 < 1,
\end{equation*}
where the first term minimizes the residual sum of squares between observations on the response variable and exogenous variables, which consist of explanatory variables $\tilde{\xmat{X}}$ and first-step predictions of the endogenous variable $\breve{\xmat{Y}}$. The constraint $\left\lVert \xvec{w} \right\rVert_1 < 1$ ensures that $||\hat{\xmat{W}}|| < 1$ and, thus, all assumptions regarding the spatial weighting matrix specified in Section 2 are met. The locations are only sampled from the center (excluding the edges of the random field), such that all entries of $\breve{\xmat{Y}}$ are known. The windows or the predicted values of two sampling locations $\xvec{s}_{*i}$ and $\xvec{s}_{*j}$ can potentially overlap. \cite{Biscio19} showed the consistency of subsampling-based statistics having an additive structure if the number of (overlapping) windows is going to infinity. In Figure \ref{fig:setting2}, we illustrate the locations and values that are used in the second step (i.e., the observation $Y(\xvec{s}_{*_i})$ and the IV-predicted values $\breve{Y}(\xvec{s}_{*_i}^{(1)}), \ldots, \breve{Y}(\xvec{s}_{*_i}^{(m)})$ for $i \in \{1,2\}$ sample locations).

The second term of the first and second-step estimation corresponds to the lasso that penalizes the absolute size of the coefficients in $\xvec{\theta}$ and $(\xvec{\beta, w})'$, respectively.
The nonnegative regularization or tuning parameters $\lambda_1$ and $\lambda_2$ are obtained by cross-validation to minimize the prediction error, as originally suggested by \cite{Tibshirani96}. However, because the large and small elements in the regression coefficients are equally driven to zero, the regular lasso yields excessively penalized large coefficients. Thus, the $\ell_1$ penalty additionally consists of individual weights such that the regression coefficients may be penalized individually (cf. \citealt{Zou06}). The weights of the first and second steps are given by the $l$-dimensional vector $\xvec{\psi}_1 = 1/|\xvec{\hat{\theta}}_0|^\gamma$, the $k$-dimensional vector $\xvec{\psi}_{2,\beta} = 1/|\xvec{\hat{\beta}}_0|^\gamma$ and the $m$-dimensional vector $\xvec{\psi}_{2,w} = 1/|\xvec{\hat{w}}_0|^\gamma$, where $\gamma > 0$. The parameter estimates $\xvec{\hat{\theta}}_0$, $\xvec{\hat{\beta}}_0$, and $\hat{\xvec{w}}_0$ can be obtained from consistent prior estimation, such as ridge regression or ordinary least squares, provided that the number of parameters is smaller than the number of observations.

\section{Simulation Study}\label{sec:MC}
In the following subsections, we analyze whether the spatial dependence structure can be recovered using the aforementioned procedure. More precisely, we critically examine the performance of the estimators with respect to (a) the number of cross-sectional resampling replications $r$, (b) the number of potential neighbors $m$, and (c) the number of the true neighbors $q$ that reflect the extent of the sparsity of the spatial weights matrix. Moreover, we analyze the performance of the approach for two distinct types of spatial dependence, namely isotropic and anisotropic processes.

\subsection{Considered settings}

For both specifications of the spatial dependence, we consider a $25 \times 25$ regular lattice with $n = 625$ grid cells, and we perform 1000 Monte Carlo replications. Regarding the spatial structure of these units, two different weighting schemes are employed. The first specification corresponds to an isotropic spatial process (Case A), where each location is equally affected by its $q = 8$ nearest neighbors with which it either shares a common edge or a vertex:
\begin{equation}
w_{ij} =
\begin{cases}
1/8 \cdot c  & \text{if } j \in N^*_i(q = 8) \land i \neq j\\
0 & \text{else}.
\end{cases}
\end{equation}
In contrast, the second specification corresponds to an anisotropic spatial process (Case B) with each location being equally affected by its neighbors to the east and south-east:
\begin{equation}
w_{ij} =
\begin{cases}
1/2 \cdot c  & \text{if } j \in N^*_i(q = 2) \land i \neq j\\
0 & \text{else}.
\end{cases}
\end{equation}
Moreover, both matrices are row-standardized such that each row sums to a constant $c$, reflecting the strength of the spatial dependence, which is consequently constant over all locations. Different degrees of positive spatial autoregressive dependence are considered, namely, $c \in \{0.5,0.7,0.9\}$.

In addition, we consider two varied sizes $m$ to determine the neighborhood structure using cross-sectional resampling, namely, $m \in \{24,48\}$. The number of replications $r$ is thereby successively increased. The minimum number of replications $r_{min} = 30$ matches the minimum requirement to conduct a 10-fold cross-validation with three observations per fold. The maximum number $r_{max} = (\sqrt{n} - \sqrt{(m+1)})^2$ includes all observations of the lattice, except for the edges. The maximum number of replications is smaller in the second step. The medium number corresponds to the average of the minimum and maximum numbers of replications (i.e., $r_{med} = \lfloor ( r_{min} + r_{max} )/2 \rfloor$). The exogenous regressors are drawn from the standard normal distribution (i.e., $X_{i,p} \sim \mathcal{N}(0,1)$ for $p = 1,\dots,k$). Moreover, $\boldsymbol{\beta}$ is a $k$-dimensional vector of ones, and $\epsilon_i \sim \mathcal{N}(0,\sigma^2)$ with $\sigma^2 = 1$.

\subsection{Results: Estimation performance}

\begin{figure}
\begin{tabular}{cc ccc}
\hline
\hline
          & & $c = 0.5$ & $c = 0.7$ & $c = 0.9$ \\
\hline
\hline\\[-.3cm]
                                    & $r_{min}$ & \begin{minipage}{0.3\textwidth}\includegraphics[width=\textwidth]{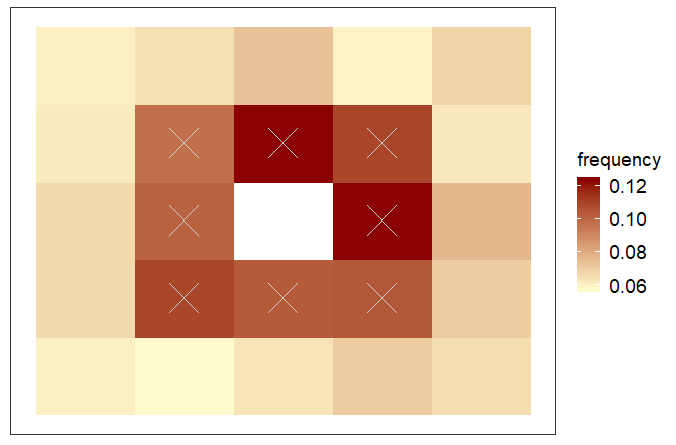}\end{minipage} & \begin{minipage}{0.3\textwidth}\includegraphics[width=\textwidth]{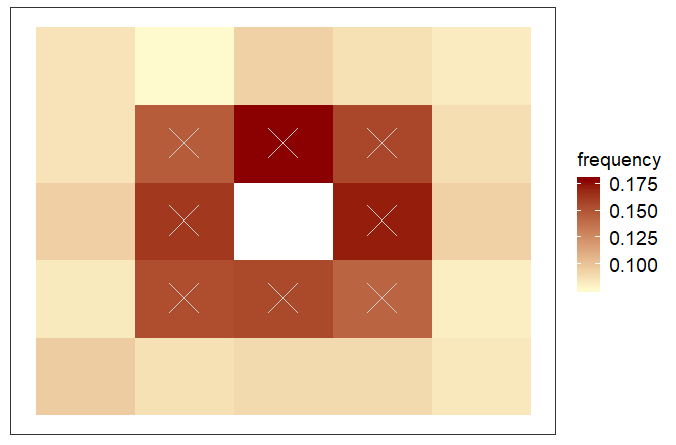}\end{minipage} & \begin{minipage}{0.3\textwidth}\includegraphics[width=\textwidth]{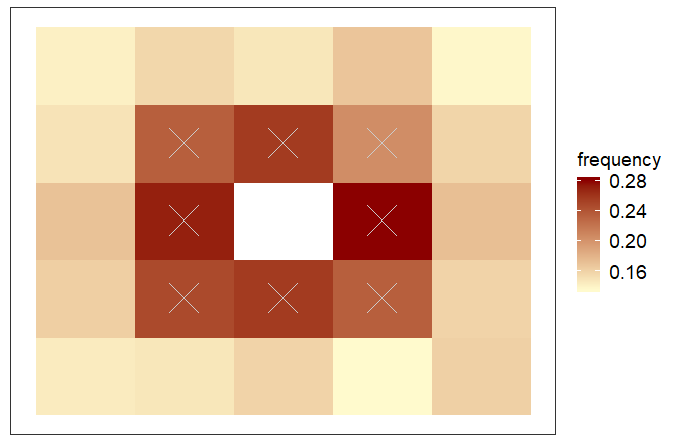}\end{minipage} \\
\rotatebox[origin=c]{90}{$m = 24$}  & $r_{med}$ & \begin{minipage}{0.3\textwidth}\includegraphics[width=\textwidth]{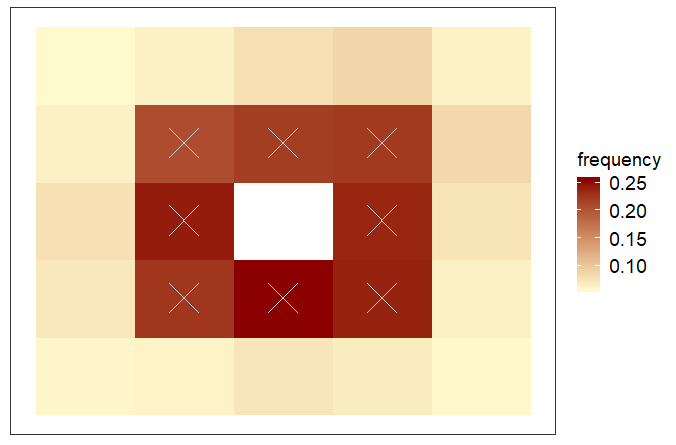}\end{minipage} & \begin{minipage}{0.3\textwidth}\includegraphics[width=\textwidth]{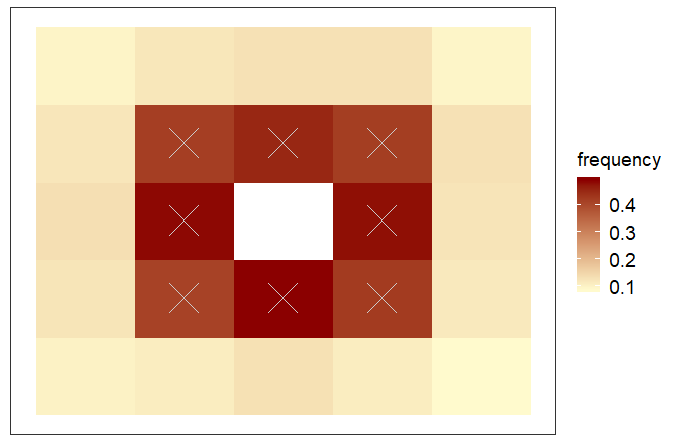}\end{minipage} & \begin{minipage}{0.3\textwidth}\includegraphics[width=\textwidth]{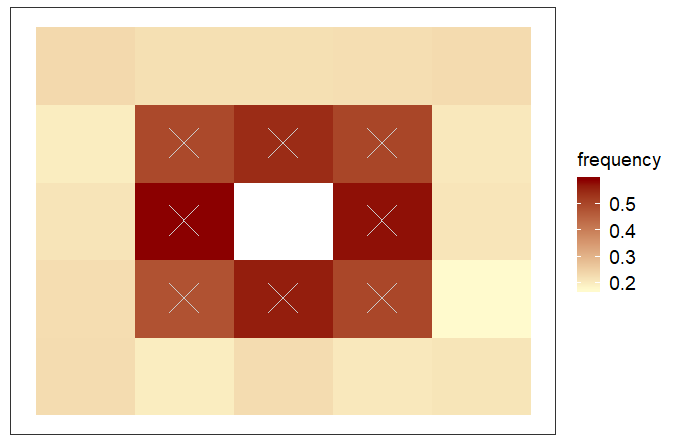}\end{minipage} \\
                                    & $r_{max}$ & \begin{minipage}{0.3\textwidth}\includegraphics[width=\textwidth]{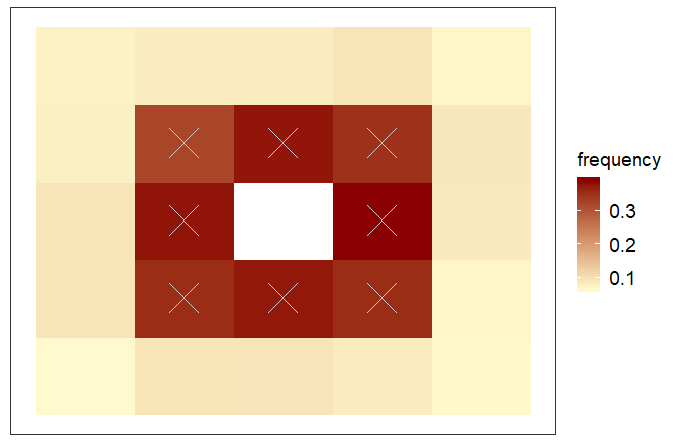}\end{minipage} & \begin{minipage}{0.3\textwidth}\includegraphics[width=\textwidth]{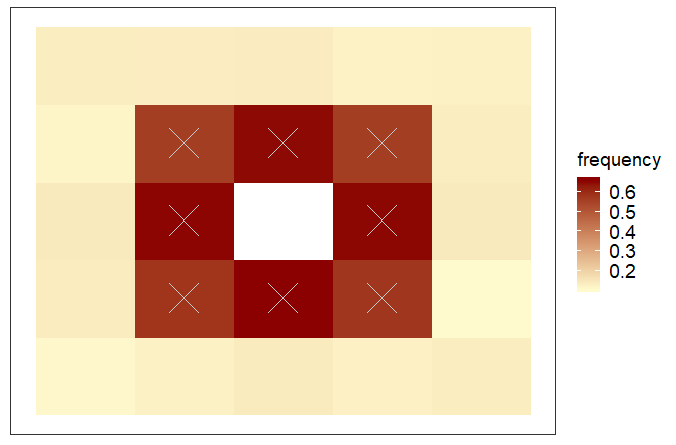}\end{minipage} & \begin{minipage}{0.3\textwidth}\includegraphics[width=\textwidth]{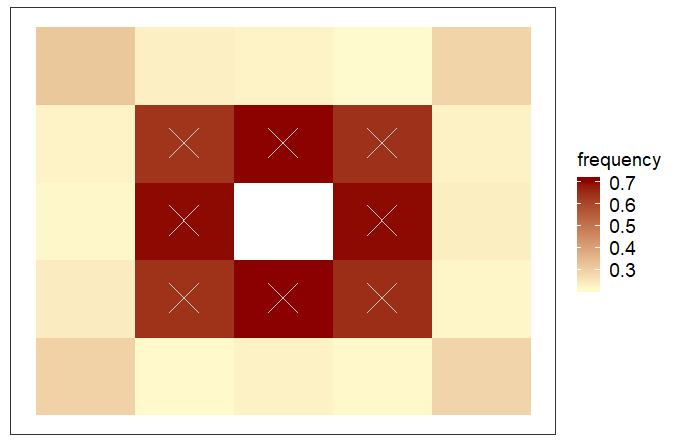}\end{minipage} \\
\hline\\[-.3cm]
                                    & $r_{min}$ & \begin{minipage}{0.3\textwidth}\includegraphics[width=\textwidth]{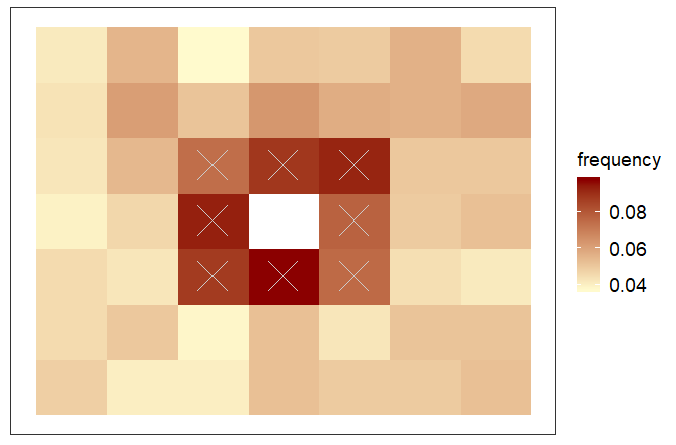}\end{minipage}   & \begin{minipage}{0.3\textwidth}\includegraphics[width=\textwidth]{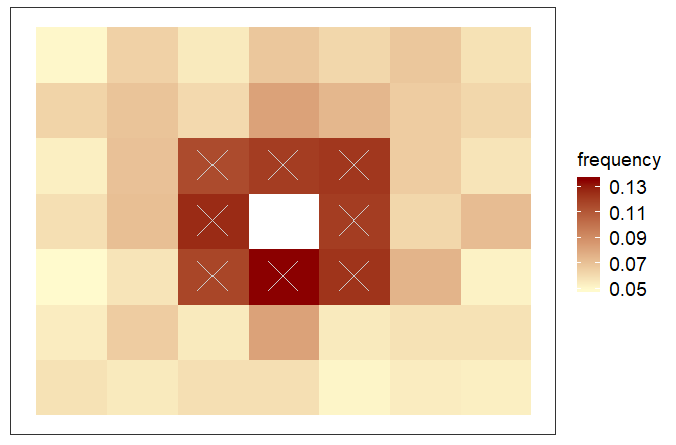}\end{minipage}   & \begin{minipage}{0.3\textwidth}\includegraphics[width=\textwidth]{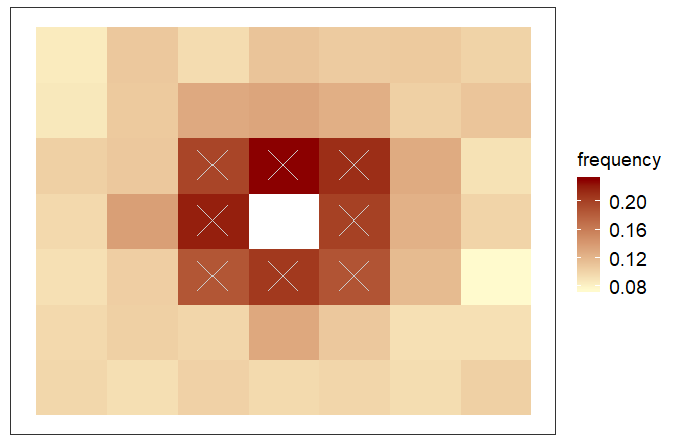}\end{minipage}   \\
\rotatebox[origin=c]{90}{$m = 48$}  & $r_{med}$ & \begin{minipage}{0.3\textwidth}\includegraphics[width=\textwidth]{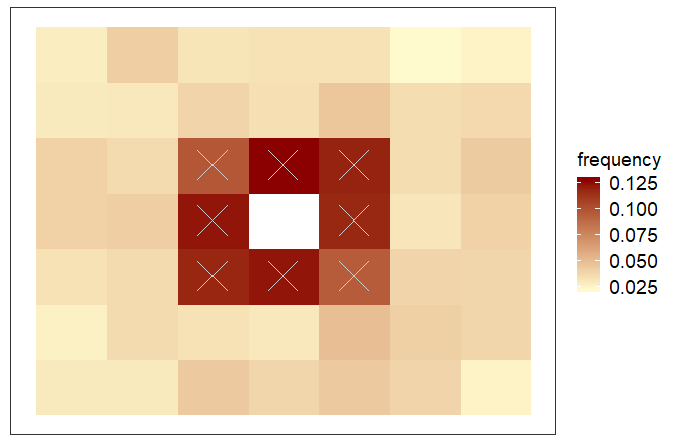}\end{minipage}   & \begin{minipage}{0.3\textwidth}\includegraphics[width=\textwidth]{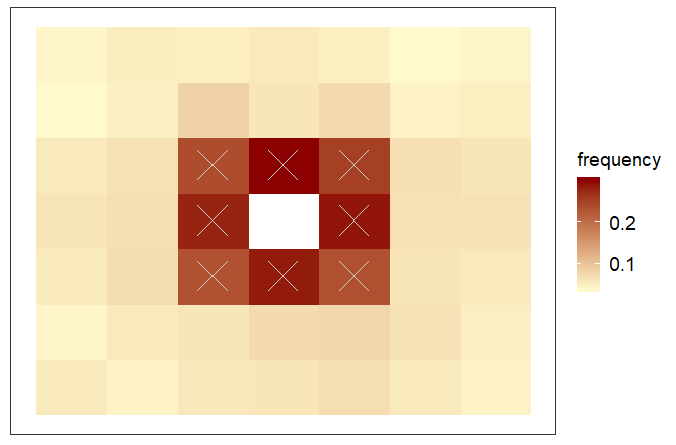}\end{minipage}   & \begin{minipage}{0.3\textwidth}\includegraphics[width=\textwidth]{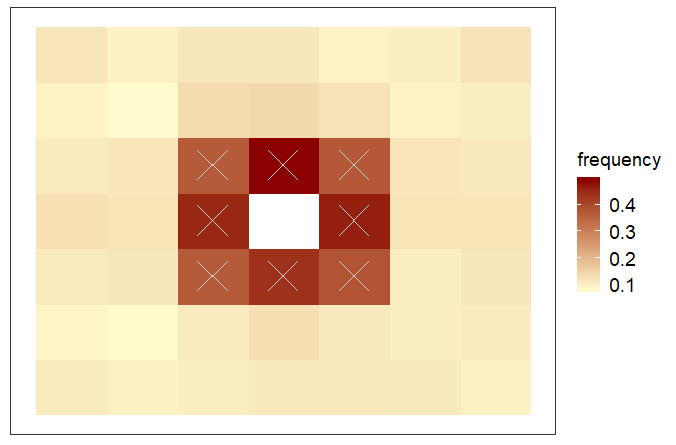}\end{minipage}   \\
                                    & $r_{max}$ & \begin{minipage}{0.3\textwidth}\includegraphics[width=\textwidth]{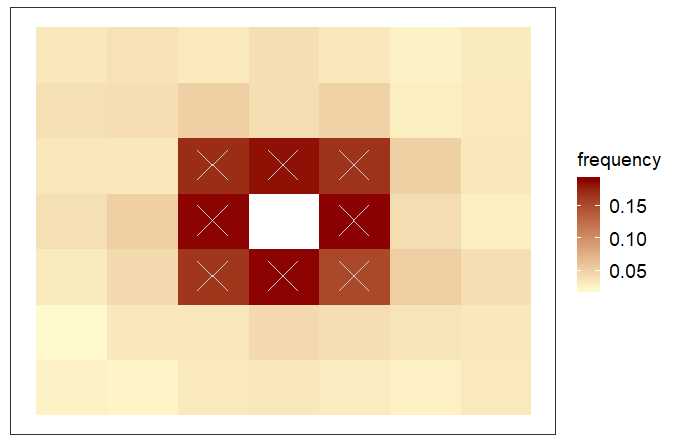}\end{minipage}   & \begin{minipage}{0.3\textwidth}\includegraphics[width=\textwidth]{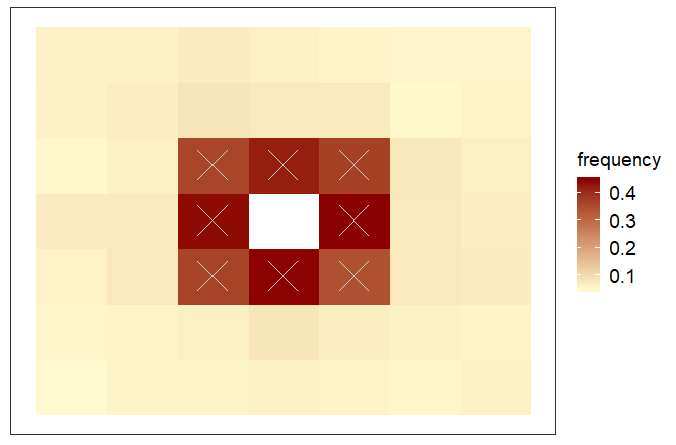}\end{minipage}   & \begin{minipage}{0.3\textwidth}\includegraphics[width=\textwidth]{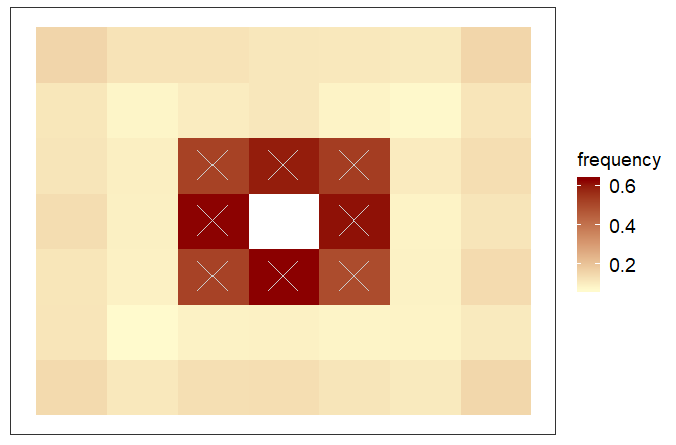}\end{minipage}  \\
\hline
\hline
\end{tabular}
\caption[Recovery frequency of isotropic $q = 8$ spatial dependence structure (Case A)]{Recovery frequency of isotropic $q = 8$ spatial dependence structure (Case A). Columns correspond to the strength of the spatial dependence $c\in \{0.5,0.7,0.9\}$. Rows correspond to the number of replications $r\in \{r_{min}, r_{med}, r_{max}\}$ for $m = 24$ and $m = 48$, respectively.
}\label{fig:Ifreq}
\end{figure}

Figure \ref{fig:Ifreq} illustrates how often each connection is identified as being non-zero for the isotropic setting (Case A). In general, the proportion of correctly identified neighbors or spatial connections increases with the number of replications $r$ and the strength of the spatial dependence $c$. Moreover, the recovery frequency is higher if only 24 instead of 48 nearest locations are considered. The true connections that are horizontally or vertically located from the center $i$ tend to be selected more often than the diagonal neighbors. This might be due to spatial spillover effects that occur because the eight first-order neighbors of $\xvec{s}_i$ are themselves influenced by their eight nearest neighbors, the second-order neighbors of $\xvec{s}_i$, and so on. Thus, the true neighbors also transmit spillover effects from higher-order neighbors that decrease strictly monotonically with increasing order. For example, if $m = 24$, the four nearest neighbors that share a common edge with $\xvec{s}_i$ possess the same number of second-order neighbors as the diagonal connections (i.e., eight), but their spillovers from the third-order neighbors are higher.

\begin{figure}
\begin{tabular}{cc ccc}
\hline
\hline
          & & $c = 0.5$ & $c = 0.7$ & $c = 0.9$ \\
\hline
\hline\\[-.3cm]
                                    & $r_{min}$ & \begin{minipage}{0.3\textwidth}\includegraphics[width=\textwidth]{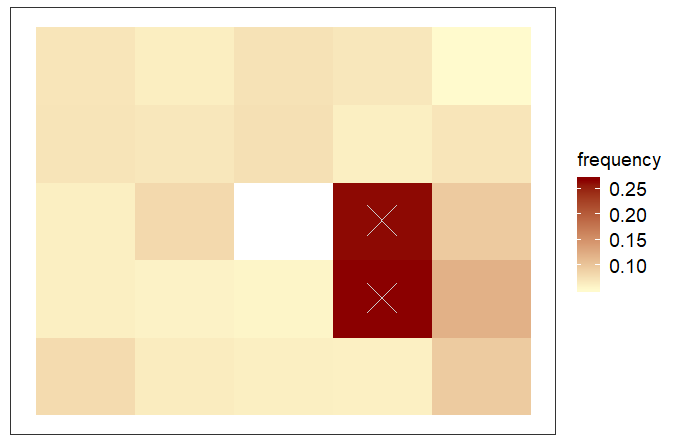}\end{minipage} & \begin{minipage}{0.3\textwidth}\includegraphics[width=\textwidth]{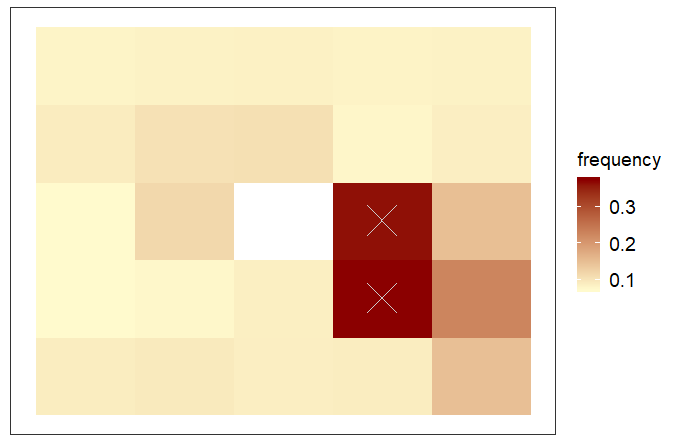}\end{minipage} & \begin{minipage}{0.3\textwidth}\includegraphics[width=\textwidth]{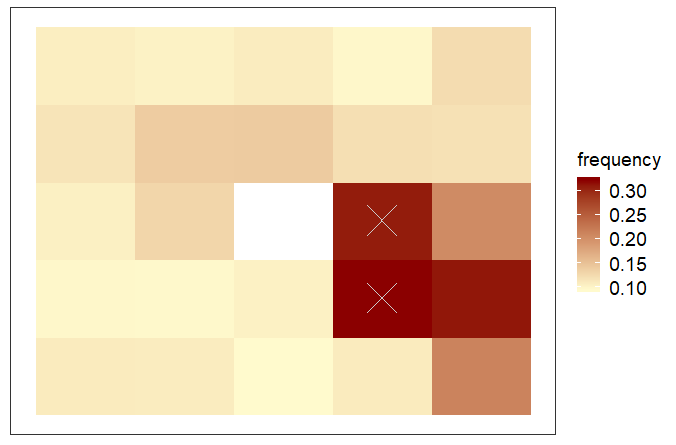}\end{minipage} \\
\rotatebox[origin=c]{90}{$m = 24$}  & $r_{med}$ & \begin{minipage}{0.3\textwidth}\includegraphics[width=\textwidth]{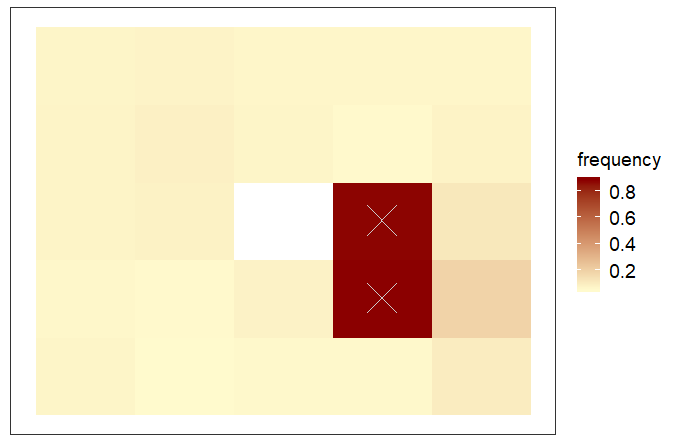}\end{minipage} & \begin{minipage}{0.3\textwidth}\includegraphics[width=\textwidth]{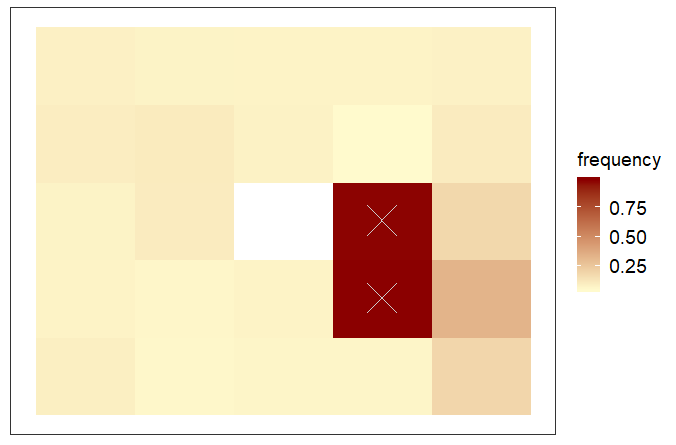}\end{minipage} & \begin{minipage}{0.3\textwidth}\includegraphics[width=\textwidth]{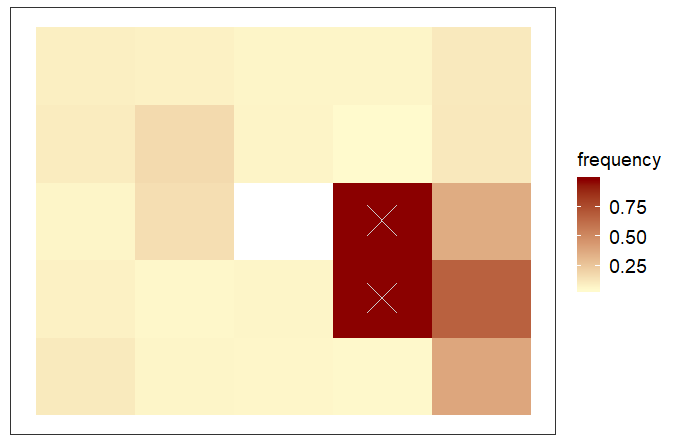}\end{minipage} \\
                                    & $r_{max}$ & \begin{minipage}{0.3\textwidth}\includegraphics[width=\textwidth]{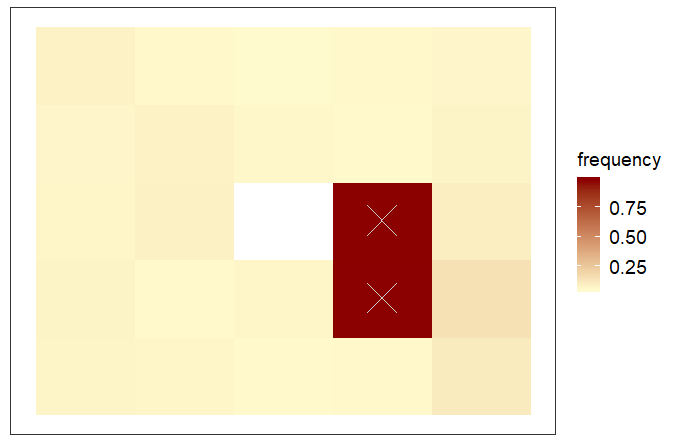}\end{minipage} & \begin{minipage}{0.3\textwidth}\includegraphics[width=\textwidth]{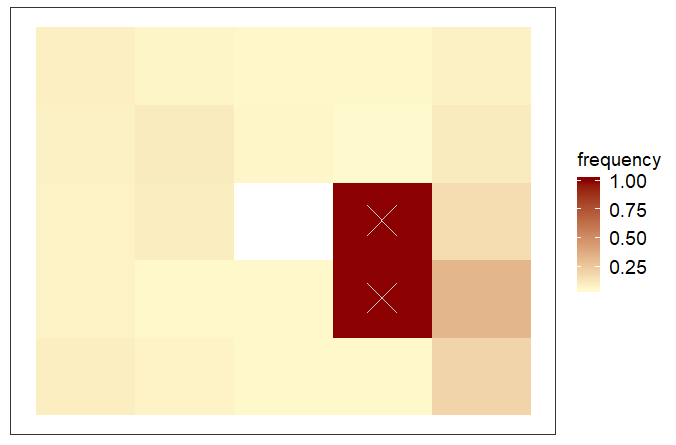}\end{minipage} & \begin{minipage}{0.3\textwidth}\includegraphics[width=\textwidth]{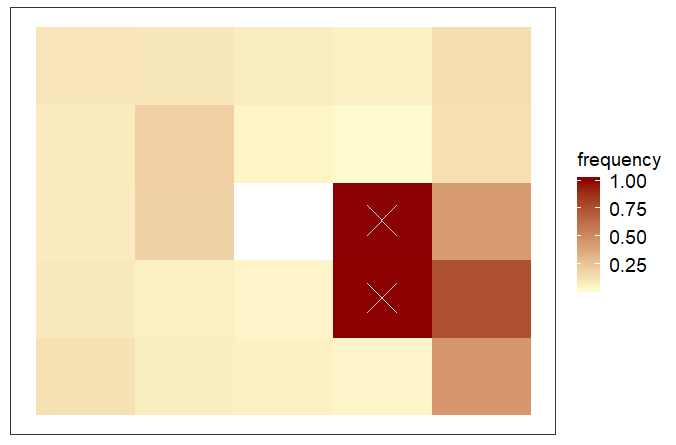}\end{minipage} \\
\hline\\[-.3cm]
                                    & $r_{min}$ & \begin{minipage}{0.3\textwidth}\includegraphics[width=\textwidth]{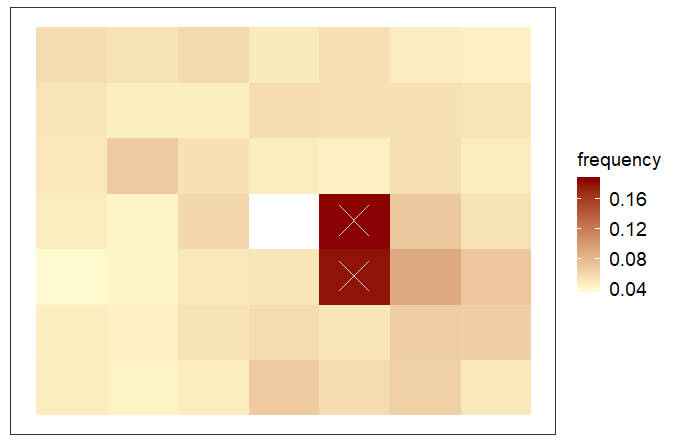}\end{minipage}   & \begin{minipage}{0.3\textwidth}\includegraphics[width=\textwidth]{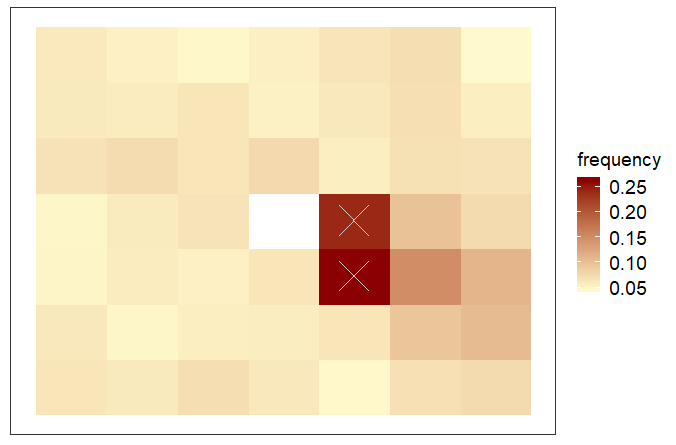}\end{minipage}   & \begin{minipage}{0.3\textwidth}\includegraphics[width=\textwidth]{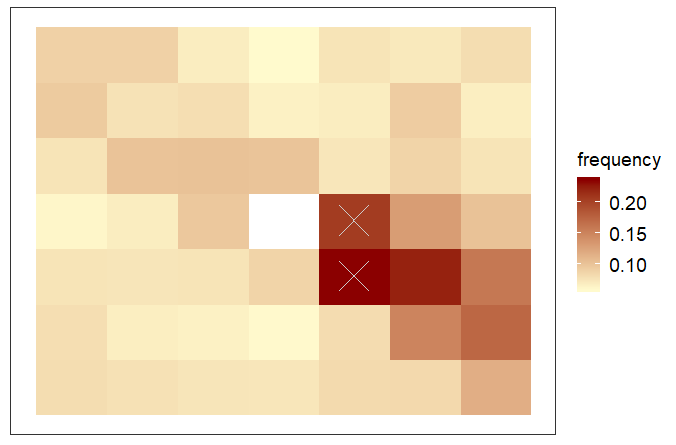}\end{minipage}   \\
\rotatebox[origin=c]{90}{$m = 48$}  & $r_{med}$ & \begin{minipage}{0.3\textwidth}\includegraphics[width=\textwidth]{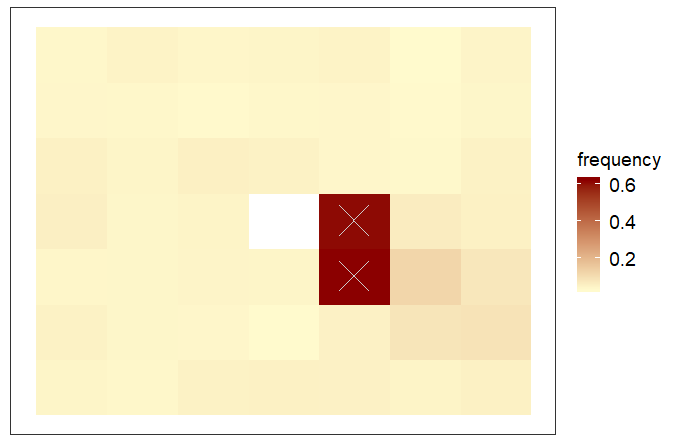}\end{minipage}   & \begin{minipage}{0.3\textwidth}\includegraphics[width=\textwidth]{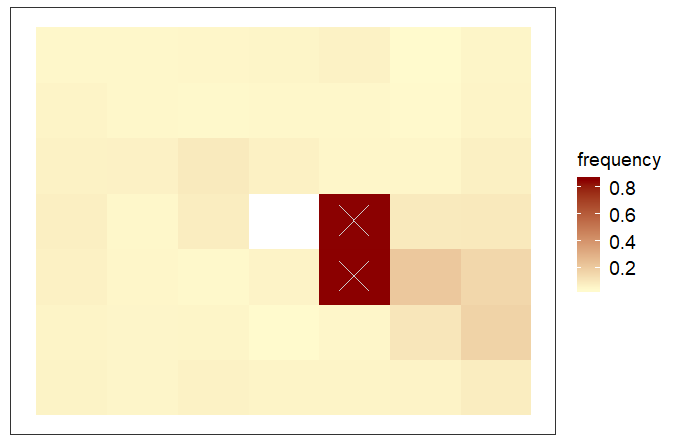}\end{minipage}   & \begin{minipage}{0.3\textwidth}\includegraphics[width=\textwidth]{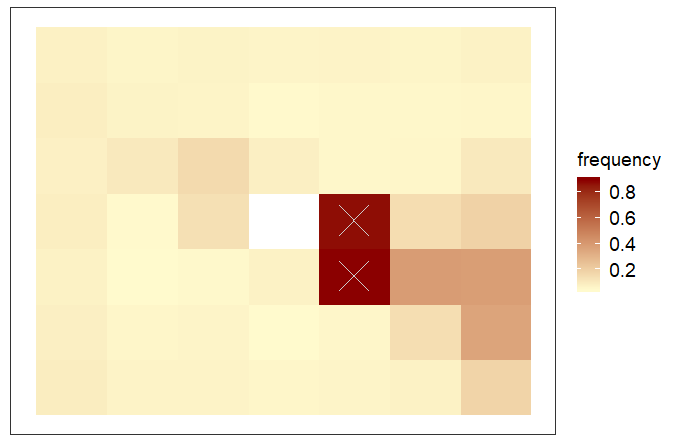}\end{minipage}   \\
                                    & $r_{max}$ & \begin{minipage}{0.3\textwidth}\includegraphics[width=\textwidth]{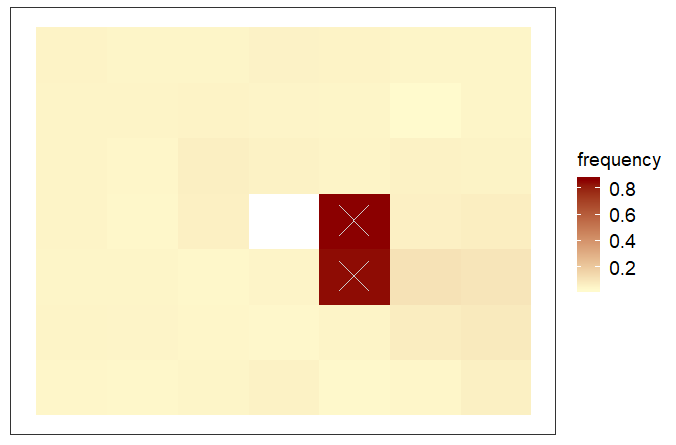}\end{minipage}   & \begin{minipage}{0.3\textwidth}\includegraphics[width=\textwidth]{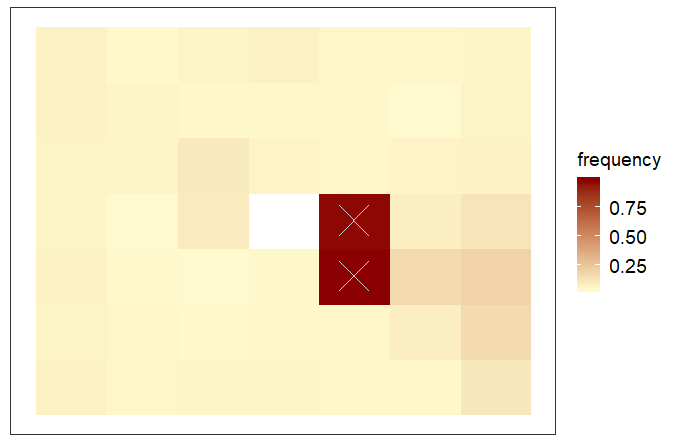}\end{minipage}   & \begin{minipage}{0.3\textwidth}\includegraphics[width=\textwidth]{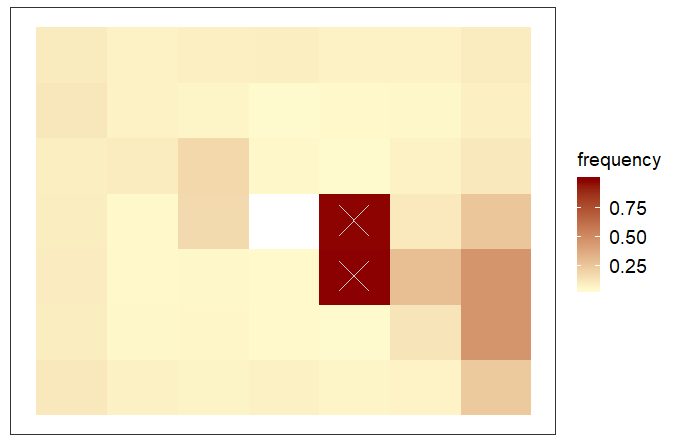}\end{minipage}  \\
\hline
\hline
\end{tabular}
\caption[Recovery frequency of anisotropic $q = 2$ spatial dependence structure (Case B)]{Recovery frequency of anisotropic $q = 2$ spatial dependence structure (Case B). Columns correspond to the strength of the spatial dependence $c\in \{0.5,0.7,0.9\}$. Rows correspond to the number of replications $r\in \{r_{min}, r_{med}, r_{max}\}$ for $m = 24$ and $m = 48$, respectively.
}\label{fig:Afreq}
\end{figure}

In contrast, Figure \ref{fig:Afreq} depicts the recovery rates for the anisotropic setting (Case B). As in the previous specification, the number of correctly identified neighbors increases with the number of replications.  However, in contrast to the first specification (isotropic setting, Case A), the recovery frequency is much higher for the correctly identified $q = 2$ non-zero connections. For instance, if $m = 24$, $c \in \{0.7,0.9\}$, and $r = r_{max}$, both true neighbors are identified across all Monte Carlo iterations. In addition, the recovery frequency is also less sensitive to the strength of the spatial dependence. Thus, stricter applications of the sparsity assumption improve the identification of the spatial dependence structure.

As in Case A, the identification also suffers from spatial spillover effects. In particular, the zero connections that are located in the east and south-east direction of the true connections are more frequently falsely selected than other zero elements if $m = 24$. Hence, the second-order neighbors are erroneously assumed to be neighbors of $i$ because they affect the center through the true connections. The effect from these second-order neighbors thereby increases with the strength of spatial dependence $c$. Increasing the number of nearest locations (i.e., $m = 48$) mitigates the influence from second-order neighbors but slightly raises the frequency of falsely selected third-order neighbors. However, the frequency of falsely selected higher-order neighbors decreases with increasing order. Moreover, the spillover effects are dispersed over several grid cells. Thus, expanding the set of nearest locations contributes to the reduction of spatial spillovers and the selection of false neighbors.

\begin{table}[!ht]
\caption[The mean absolute error (MAE) of the regression coefficient $\beta$ and $\xvec{w}$, specificity $\Pi_0$, and sensitivity $\Pi_1$ of the estimated spatial weights depending on the type of weighting scheme $q$, strength of the spatial dependence $c$, number of cross-sectional resampling replications $r$, and number of potential neighbors $m$]{The mean absolute error (MAE) of the regression coefficient $\beta$ and $\xvec{w}$, specificity $\Pi_0$, and sensitivity $\Pi_1$ of the estimated spatial weights depending on the type of weighting scheme $q$, strength of the spatial dependence $c$, number of cross-sectional resampling replications $r$, and number of potential neighbors $m$.}\label{tab:MC}
\centering
\scriptsize
\begin{tabular}{llllcccccccc}
\hline
\hline
& &    &  & $c = 0.5$ & & & $c = 0.7$ & & & $c = 0.9$ & \\
$q$ & $m$ &  & $r_{min}$ & $r_{med}$ & $r_{max}$ & $r_{min}$ & $r_{med}$ & $r_{max}$ & $r_{min}$ & $r_{med}$ & $r_{max}$\\
\hline
\hline\\[-.2cm]
\multicolumn{12}{c}{Isotropic setting (case A)} \\[.1cm]
8 & 25 & $\mathrm{MAE}_\beta$ & 0.2120 & 0.0698 & 0.0519 & 0.2592 & 0.0810 & 0.0545 & 0.4638 & 0.1026 & 0.0591 \\
 &  & $\mathrm{MAE}_w$ & 0.0453 & 0.0245 & 0.0223 & 0.0602 & 0.0336 & 0.0276 & 0.0904 & 0.0504 & 0.0403 \\
 &  & $\Pi_0$ & 0.8034 & 0.8057 & 0.7544 & 0.7877 & 0.7198 & 0.6823 & 0.7333 & 0.6206 & 0.5708 \\
& & $\Pi_1$ & 0.2248 & 0.3543 & 0.5239 & 0.2735 & 0.6148 & 0.7825 & 0.3511 & 0.7110 & 0.8561 
\vspace{0.2cm}\\
8 & 49 & $\mathrm{MAE}_\beta$ & 0.2227 & 0.0917 & 0.0699 & 0.2819 & 0.1079 & 0.0775 & 0.4649 & 0.1391 & 0.0828 \\
 & & $\mathrm{MAE}_w$ & 0.0455 & 0.0139 & 0.0131 & 0.0606 & 0.0199 & 0.0183 & 0.0472 & 0.0298 & 0.0259 \\
 &  & $\Pi_0$ & 0.8724 & 0.8786 & 0.8617 & 0.8569 & 0.8305 & 0.7880 & 0.8322 & 0.7454 & 0.7119 \\
& & $\Pi_1$ & 0.1496 & 0.2089 & 0.2886 & 0.1834 & 0.4020 & 0.5771 & 0.2450 & 0.6148 & 0.7578
\vspace{0.1cm}\\
\multicolumn{12}{c}{Anisotropic setting (case B)} \\[.1cm]
2 & 25 & $\mathrm{MAE}_\beta$ & 0.2324 & 0.0689 & 0.0487 & 0.2994 & 0.0764 & 0.0531 & 0.4859 & 0.1072 & 0.0718 \\
 & & $\mathrm{MAE}_w$ & 0.0406 & 0.0134 & 0.0091 & 0.0519 & 0.0151 & 0.0107 & 0.0735 & 0.0286 & 0.0229 \\
 & & $\Pi_0$ & 0.7926 & 0.7508 & 0.7265 & 0.7685 & 0.7054 & 0.6812 & 0.7645 & 0.6434 & 0.6057 \\
& & $\Pi_1$ & 0.3755 & 0.9345 & 0.9915 & 0.4710 & 0.9900 & 1 & 0.4035 & 0.9865 & 1 \vspace{0.2cm}\\
2 & 49 & $\mathrm{MAE}_\beta$ & 0.2600 & 0.0959 & 0.0669 & 0.3460 & 0.1066 & 0.0730 & 0.5355 & 0.1487 & 0.0964 \\
 &  & $\mathrm{MAE}_w$ & 0.0288 & 0.0112 & 0.0085 & 0.0377 & 0.0133 & 0.0097 & 0.1591 & 0.0217 & 0.0174 \\
 &  & $\Pi_0$ & 0.8624 & 0.8309 & 0.7985 & 0.8545 & 0.7784 & 0.7455 & 0.8375 & 0.7178 & 0.6697 \\
& & $\Pi_1$ & 0.2740 & 0.7350 & 0.9155 & 0.3145 & 0.9125 & 0.9905 & 0.2715 & 0.9315 & 0.9915 \\
\hline
\hline
\end{tabular}
\end{table}

To evaluate the identification of the spatial weights and the performance of the corresponding estimates, we report several statistics in Table \ref{tab:MC} for both the isotropic and anisotropic settings ($k = 1$). More precisely, specificity $\Pi_0$ refers to the average percentage of correctly identified zero elements, while sensitivity $\Pi_1$ indicates the average percentage of correctly identified non-zero weights. In addition, the mean absolute error (MAE) is calculated as a measure of dispersion between the true and estimated spatial weights.
More precisely, the MAE of the $t$-th Monte Carlo iteration is calculated as follows:
\begin{equation}
\mathrm{MAE}_{(t)} = m^{-1} \left\lVert \hat{\xvec{w}}_{(t)} - \xvec{w}_{(t)} \right\rVert_1.
\end{equation}

Increasing the number of replications $r$ results in higher sensitivity values, lower specificity values, and a lower MAE. Thus, both estimators for $\xvec{w}$ and $\xvec{\beta}$ are asymptotically consistent for the increasing sampling replications $r$. Moreover, the spatial dependence estimator is selection-consistent with increasing $r$ for both settings (Cases A and B).

However, if the number of potential neighbors $m$ is extended, the true connections are harder to determine, resulting in higher specificity and lower sensitivity values. Thus, the percentage of correctly selected neighbors increases with the number of observations and the strength of the spatial dependence $c$, matching the increase in the individual recovery frequencies, but it decreases with $m$.

The frequency of correctly selected neighbors does not exceed $86\%$ for Case A, implying that, on average, at least one out of eight neighbors is not identified. At the same time, the proportion of falsely selected zero elements increases with $r$. This is in line with the findings by \cite{Ahrens15}, where the average percentage of zero elements that are falsely selected as being non-zero increases with the number of time periods. The proportion of zero elements that are erroneously selected as being non-zero maintains a relatively high level. \cite{Buhlmann11} pointed out that false-positive rates (i.e., falsely selected variables) can typically not be avoided due to the lasso's property to select too many variables. Moreover, high values of $\Pi_0$ do not necessarily indicate good model performance because setting all weights to zero yields a value of $\Pi_0 = 1$.

Higher individual recovery frequencies are also reflected in the overall sensitivity values, which are much higher in the case of anisotropic dependencies (Case B). In particular, for $c \in \{0.7, 0.9\}$ and $r = r_{max}$, the average percentage of correctly identified non-zero connections is approximately $100\%$. As before, the average percentage of falsely selected zero connections increases with $r$. In particular, the specificity values are remarkably similar to those of the first specification. Therefore, while the extent of the sparsity has an influence on the selection of the true neighbors, it does not substantially affect the identification of zero connections.

\subsection{Results: Computation time}
\begin{figure}
\centering
\includegraphics[scale=0.9]{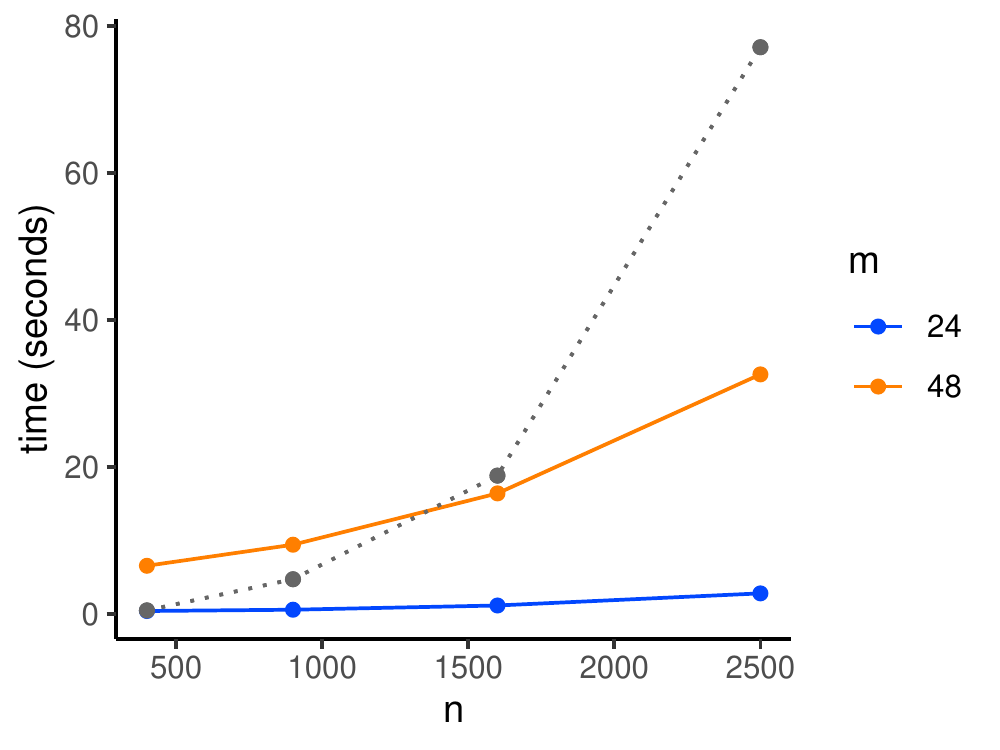}
\caption[Average computation time]{Average computation time of the two-step adaptive lasso approach with $n \in \{400,900,1600,2500\}$ and $r = r_{max}$ for $m = 24$ (blue), $m = 48$ (orange) and the average computation time of the maximum likelihood estimation (gray, dotted) with deterministic weights as the benchmark approach.}\label{fig:comptime}
\end{figure}

Finally, Figure \ref{fig:comptime} depicts the average computation time per iteration required to conduct the two-step adaptive lasso approach, where the maximum number of replications $r_{max}$ is considered for cross-sectional resampling. Accordingly, the computational complexity depends not only on the sample size $n$ but also on the size of the set of potential neighbors $m$. For comparison, the computation time resulting from the model estimation using a classical ML approach with deterministic weights is illustrated. While no considerable difference exists in computation time for smaller sample sizes, the two-step lasso offers clear advantages for larger sample sizes of $n > 1600$. This is primarily due to the circumvention of complex matrix operations, such as inversion or computation of log-determinants associated with ML estimation.

To summarize, the number of cross-sectional resampling replications is crucial for the identification of the spatial weights structure and the performance of the corresponding estimates. In particular, the average percentage of correctly identified non-zero weights and the recovery frequency of each individual neighbor increase. At the same time, the estimation accuracy (in terms of the MAE) improves with the number of replications. The specificity statistic $\Pi_0$ or the number of correctly identified zero connections alternatively decreases if $r$ is increased. This natural effect results from spatial spillover effects that lead to false selections of higher-order neighbors and the lasso's tendency to select too many variables. By increasing the number of potential neighbors $m$, the individual effects from these false neighbors can be dispersed and diminished. However, if $m = 48$, the individual neighbors are more difficult to determine. Moreover, an expansion of the set of potential neighbors always coincides with a reduction in the cross-sectional resampling replications, which is especially relevant for smaller sample sizes.
Regarding the identifying assumption of the sparsity of the spatial weights matrix, which is reflected by the number of true neighbors $q$, sparse matrices with few non-zero weights are better identified than dependence structures that are more connected.

\section{Empirical Application}
In this section, we apply the proposed approach to model $\mathrm{NO_2}$ pollution in parts per billion (ppb) for Mexico. We consider the satellite-derived three-year mean ground-level concentrations for the years 2000 to 2002 with a grid-cell resolution of 0.1$^{\circ}$, covering a land surface from 21$^\circ$ to 27$^\circ$ north and 99$^\circ$ to 105$^\circ$ west, which yields $n = 3540$ observations. Both $\mathrm{NO_{2}}$ concentrations and the exogenous regressors, namely, $\mathrm{PM_{2.5}}$ concentrations, temperature, population count data, and elevation levels, are retrieved from the NASA Socioeconomic Data and Applications Center.

To investigate the spatial neighborhood structure, the influence of the $m = 48$ nearest locations is examined in the first and second steps, respectively. To assess whether the estimation of the spatial dependence structure improves the model performance, we consider two additional competing model specifications. In particular, two deterministic, row-normalized weighting schemes are assumed: a queen matrix with $q = 8$ nearest neighbors and a rook matrix with $q = 4$ nearest neighbors to the north, south, east, and west. For inference and better comparison, we additionally estimate the two deterministic model specifications using the ML approach. 

To obtain standard errors of the lasso estimators, a bootstrap with 100 iterations and $r = 600$ randomly selected locations is employed. Hence, both standard errors and regression coefficients are averaged over the bootstrap iterations. However, concerning the significance of these coefficients, standard errors are not meaningful in the context of strongly penalized and, therefore, biased coefficients (see \citealt{Goeman18}). Eventually, root mean square errors (RMSEs) are calculated as a measure of dispersion between the estimated and observed $\mathrm{NO_2}$ values:
\begin{equation}
\mathrm{RMSE} = n^{-\frac{1}{2}} \left\lVert \hat{\xvec{Y}} - \xvec{Y} \right\rVert_2,
\end{equation}
where $\hat{\xvec{Y}}$ denotes the estimated values that are obtained using either the second-step coefficients or the ML estimates, depending on the respective estimation procedure.

Thus, we compare five different models in total. First, we estimate the spatial dependence structure using the bootstrap-based two-step lasso approach (Model 1). Furthermore, both queen and rook specifications are estimated using the same lasso procedure (Models 2 and 3) and the ML approach (Models 4 and 5).
Table \ref{reg} reports the parameter estimates and RMSE for all models.

\begin{table}[!ht]
\caption[Parameter estimates, standard errors in parentheses, and root mean square errors (RMSEs) for the (1) two-step lasso with estimated spatial weights; (2) two-step lasso with the queen matrix; (3) two-step lasso with the rook matrix; (4) maximum likelihood (ML) with the queen matrix; and (5) maximum likelihood (ML) with the rook matrix]{Parameter estimates, standard errors in parentheses, and root mean square errors (RMSEs) for the (1) two-step lasso with estimated spatial weights; (2) two-step lasso with the queen matrix; (3) two-step lasso with the rook matrix; (4) maximum likelihood (ML) with the queen matrix; and (5) maximum likelihood (ML) with the rook matrix.}\label{reg}
\centering
\begin{tabular}{lcccccc}
\hline
\hline
  &  (1) & (2) & (3) &  (4) &  (5) \\
\hline
\hline
$c$ & 0.9254 & 0.9999 & 0.9997 & 0.9942 & 0.9955 \\
 & (0.0257) & (0.0004)  & (0.0016) &  (0.0013) & (0.0012)  \\
$\beta_1$ & 0.0100 & -0.0240 & -0.0203 & -0.0005 & -0.0005  \\
\tiny{($\mathrm{PM_{2.5}}$)} & (0.0236) & (0.0276)  & (0.0304) &  (0.0040) & (0.0032) \\
$\beta_2$ &  -0.0077 & -0.0043 &  -0.0127 &  0.0044 & 0.0026 \\
\tiny{(temperature)} & (0.0274) & (0.0338) & (0.0331) &  (0.0044) & (0.0035) \\
$\beta_3$ &  0.1048 & 0.0716 & 0.0585 & 0.0785 & 0.0562 \\
\tiny{(population)} & (0.0882) & (0.0864)  & (0.0980) &  (0.0031) & (0.0025) \\
$\beta_4$ &  -0.0666 & -0.0911 &  -0.1007 & -0.0040 & -0.0041  \\
\tiny{(elevation)} & (0.0781) & (0.0418) & (0.0480) &  (0.0048) & (0.0040) \\
\vspace{-0.2cm}\\
RMSE & 0.0895 & 0.1577 &  0.1718 &  0.3373 & 0.2571 \\
\hline
\hline
\end{tabular}\\
\raggedright
\footnotesize{Standard errors and parameter estimates are obtained by bootstrapping for Models 1-3 and as Cramer-Rao bound for Models 4-5. Variables are standardized.}
\end{table}                  
                  
\begin{figure}
\centering
\includegraphics[width=0.8\textwidth]{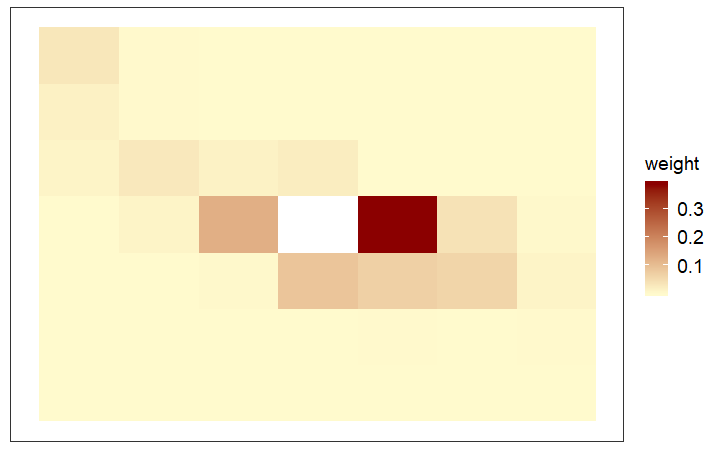}
\caption[Estimated spatial dependence structure for $\mathrm{NO_2}$ concentrations with $\sum\limits_{i=1}^m=0.9254$ and $m = 48$ potential neighbors]{Estimated spatial dependence structure for $\mathrm{NO_2}$ concentrations with $\sum\limits_{i=1}^m=0.9254$ and $m = 48$ potential neighbors.}\label{fig:weights}
\end{figure}

Regardless of the model specification, the parameter estimates $\hat{c}$ indicate very strong spatial dependencies between $\mathrm{NO_2}$ concentrations. If the spatial weights are estimated, the sum of the individual weights in $\hat{\xvec{w}}$ indicates the strength of that spatial dependence, namely, $\hat{c} = 0.9254$. Figure \ref{fig:weights} illustrates the corresponding estimated spatial dependence structure. Accordingly, the individual locations are primarily affected by their adjacent neighbor to the east and, to a lesser extent, by their neighbors to the west and south. In addition, some higher-order neighbors to the east and north-west are of minor importance. Hence, the resulting dependence structure is anisotropic and highly irregular, resembling neither a queen nor rook specification.
In contrast, when employing a prespecified queen or rook matrix, the spatial autoregressive parameter is directly estimated. Both the two-step adaptive lasso and the ML approach yield parameter estimates that are very close to 1, which marks the upper bound of the feasible parameter space, assuming row-normalized spatial weights.

Concerning the effect of exogenous variables, the competing model specifications lead to different results because the regression coefficients depend on the choice of both the weighting matrix and estimation procedure. In general, the discrepancies are greater for the estimation approach than for the choice of the weighting matrix. Regarding the effect of the population size and elevation, all model specifications are consistent regarding the sign of the regression coefficients $\beta_3$ and $\beta_4$, respectively. Accordingly, the population size has a positive effect on $\mathrm{NO_2}$, while the elevation has a negative effect (see, e.g., \citealt{Skene10}).

\begin{figure}
\begin{tabular}{c ccc}
\hline
\hline
          & 21$^\circ$ to 27$^\circ$ north & 27$^\circ$ to 30$^\circ$ north   & 40$^\circ$ to 45$^\circ$ north  \\
          & 99$^\circ$ to 105$^\circ$ west & 97.5$^\circ$ to 105$^\circ$ west & 114$^\circ$ to 120$^\circ$ west \\
\hline
\hline
\multicolumn{4}{c}{$\mathrm{NO_2}$ observations} \\[-.1cm]
\rotatebox[origin=c]{90}{Observations}    & \begin{minipage}{0.3\textwidth}\includegraphics[width=\textwidth]{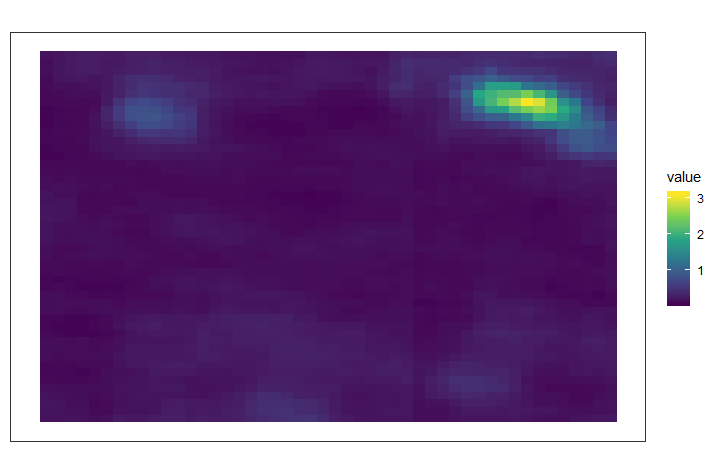}\end{minipage}     & \begin{minipage}{0.3\textwidth}\includegraphics[width=\textwidth]{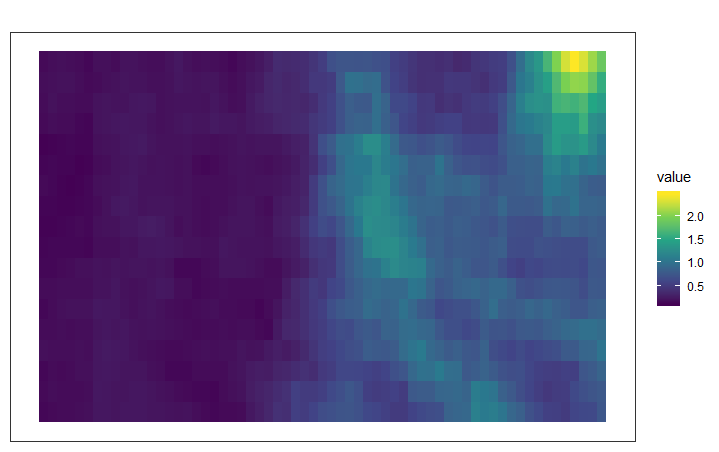}\end{minipage}     & \begin{minipage}{0.3\textwidth}\includegraphics[width=\textwidth]{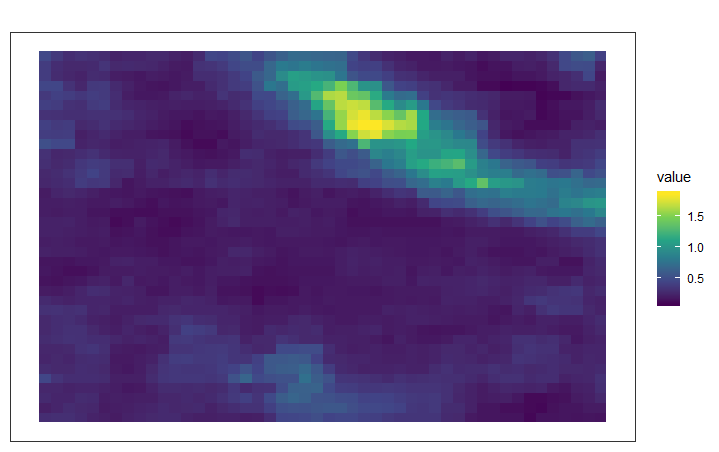}\end{minipage} \\
\hline
\multicolumn{4}{c}{Two-step lasso} \\[-.1cm]
\rotatebox[origin=c]{90}{Estimated}       & \begin{minipage}{0.3\textwidth}\includegraphics[width=\textwidth]{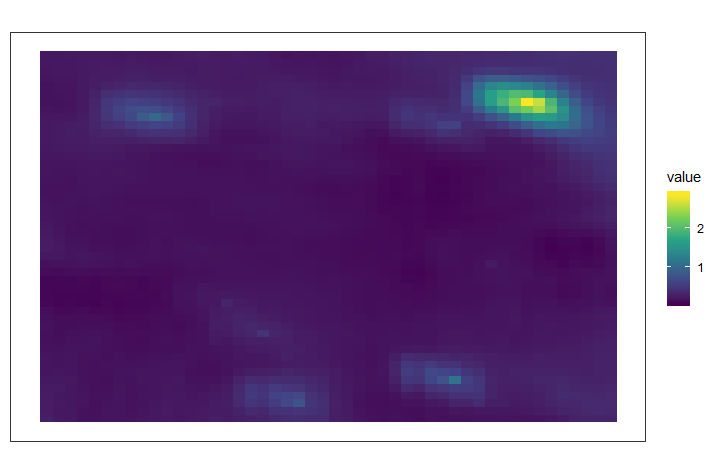}\end{minipage}      & \begin{minipage}{0.3\textwidth}\includegraphics[width=\textwidth]{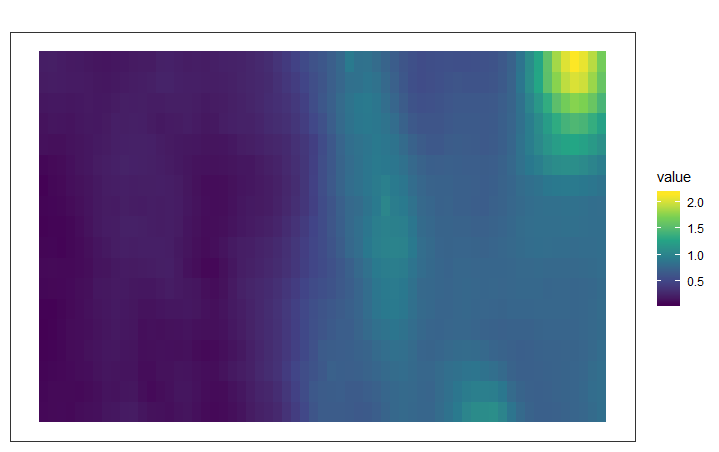}\end{minipage}      & \begin{minipage}{0.3\textwidth}\includegraphics[width=\textwidth]{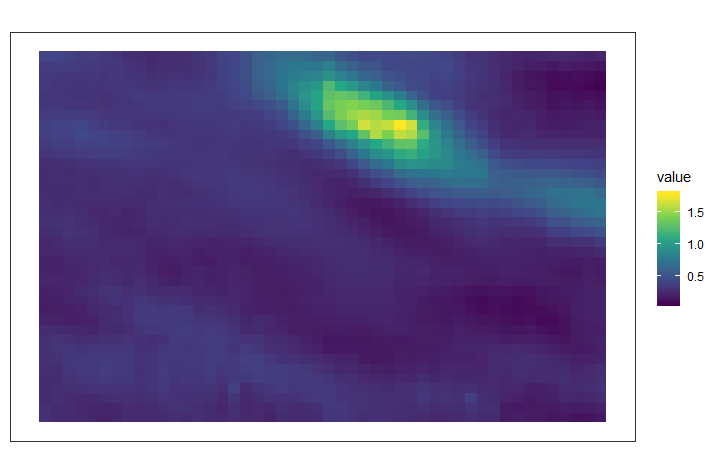}\end{minipage} \\
\rotatebox[origin=c]{90}{Queen}  & \begin{minipage}{0.3\textwidth}\includegraphics[width=\textwidth]{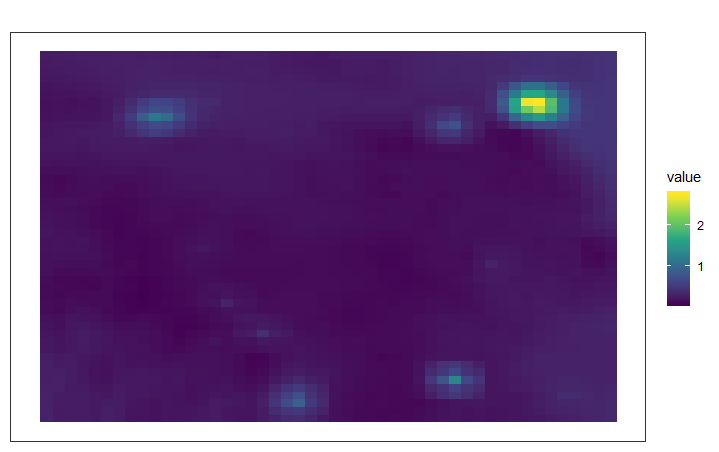}\end{minipage} & \begin{minipage}{0.3\textwidth}\includegraphics[width=\textwidth]{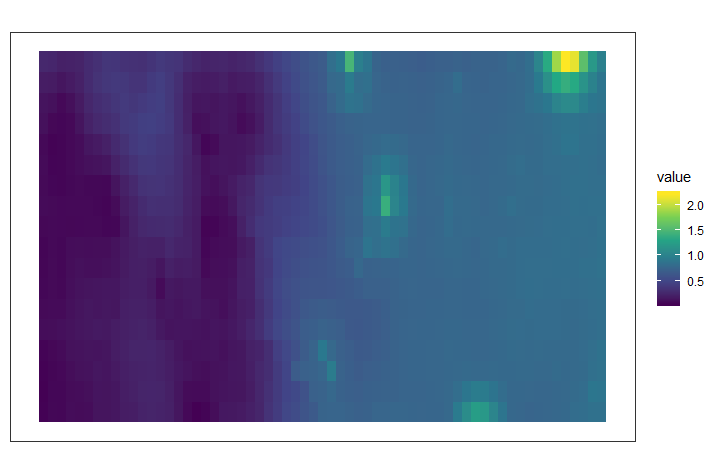}\end{minipage} & \begin{minipage}{0.3\textwidth}\includegraphics[width=\textwidth]{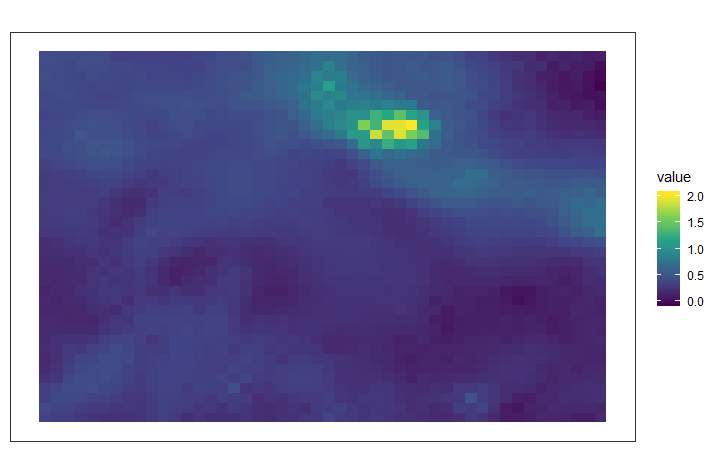}\end{minipage} \\
\rotatebox[origin=c]{90}{Rook}        & \begin{minipage}{0.3\textwidth}\includegraphics[width=\textwidth]{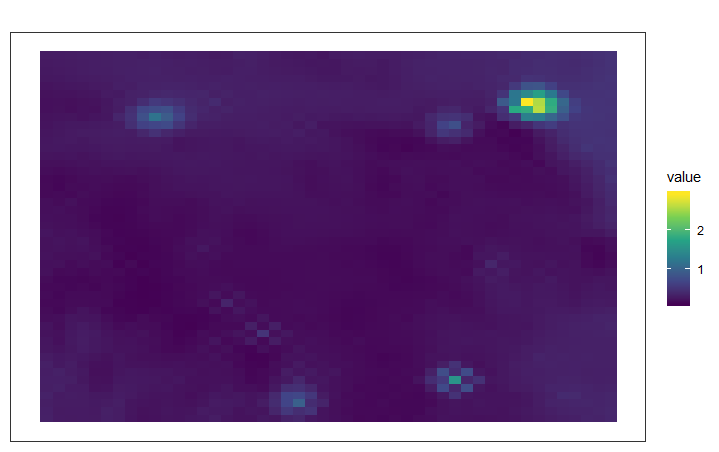}\end{minipage}  & \begin{minipage}{0.3\textwidth}\includegraphics[width=\textwidth]{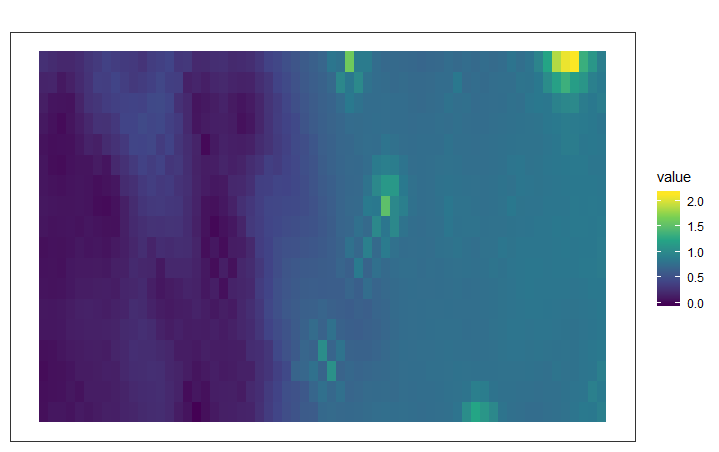}\end{minipage}  & \begin{minipage}{0.3\textwidth}\includegraphics[width=\textwidth]{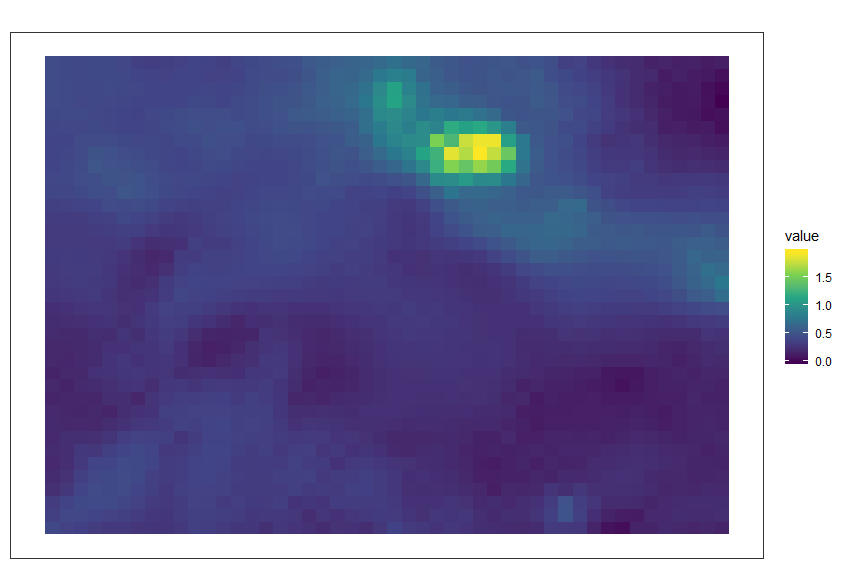}\end{minipage}   \\
\hline
\multicolumn{4}{c}{ML estimation} \\[-.1cm]
\rotatebox[origin=c]{90}{Queen}   & \begin{minipage}{0.3\textwidth}\includegraphics[width=\textwidth]{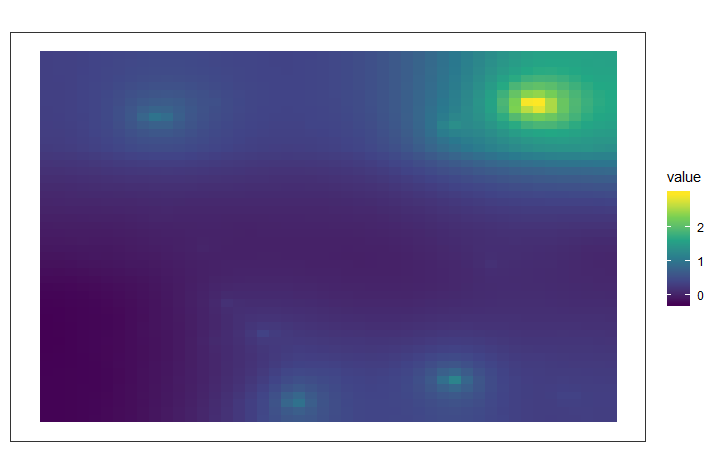}\end{minipage}  & \begin{minipage}{0.3\textwidth}\includegraphics[width=\textwidth]{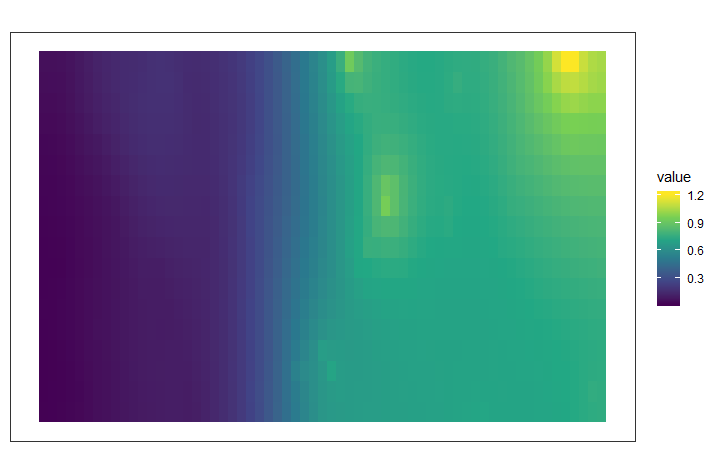}\end{minipage}  & \begin{minipage}{0.3\textwidth}\includegraphics[width=\textwidth]{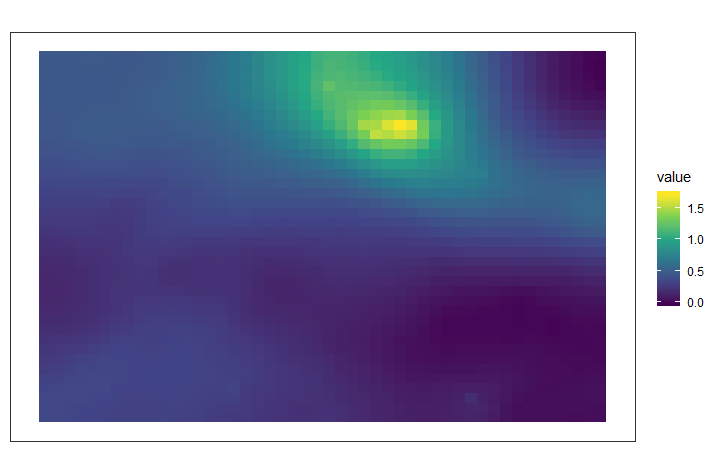}\end{minipage}   \\
\rotatebox[origin=c]{90}{Rook}   & \begin{minipage}{0.3\textwidth}\includegraphics[width=\textwidth]{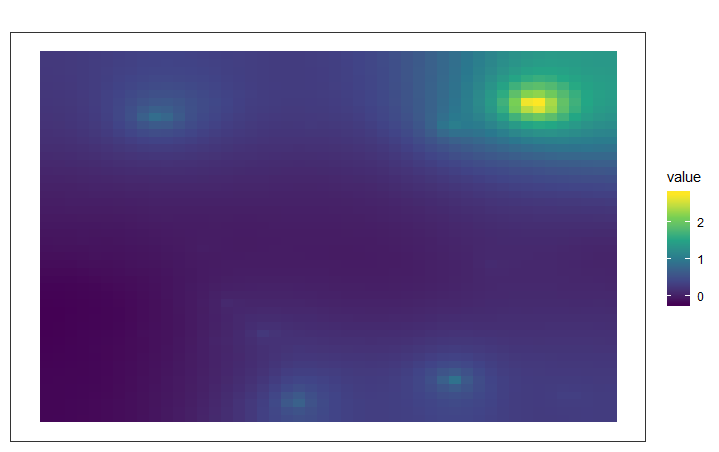}\end{minipage}   & \begin{minipage}{0.3\textwidth}\includegraphics[width=\textwidth]{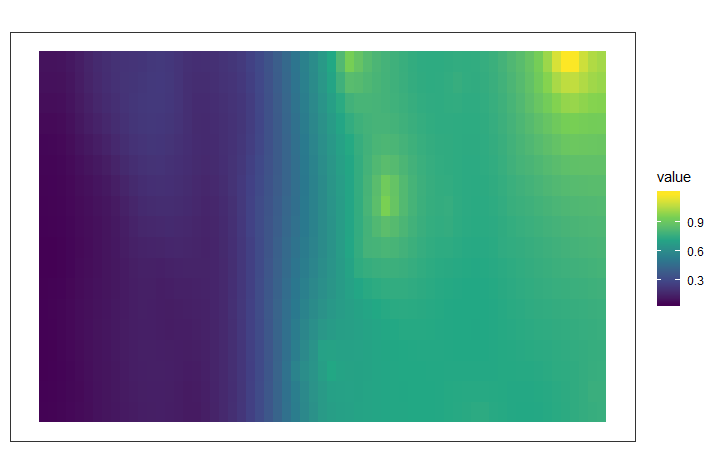}\end{minipage}   & \begin{minipage}{0.3\textwidth}\includegraphics[width=\textwidth]{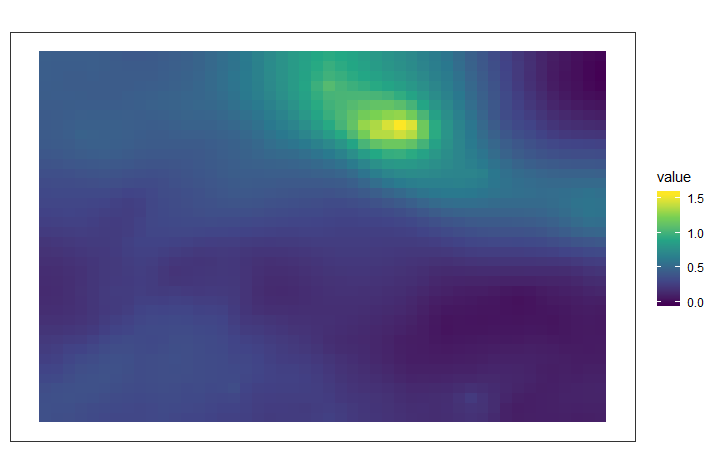}\end{minipage}  \\
\hline
\hline
\end{tabular}
\thispagestyle{empty}
\caption[Empirical examples of the model fit]{Empirical examples of the model fit. The $\mathrm{NO_2}$ observations at selected locations ($1^{st}$ row) and the model fit of the two-step lasso approach with the estimated spatial weights, queen's matrix, rook's matrix ($2^{nd}$-$4^{th}$ row), and ML approach as the benchmark approach ($5^{th}$-$6^{th}$ row).}\label{fig:pred}
\end{figure}

Finally, estimating the spatial weights structure leads to an RMSE reduction of around 43\% and 48\% in contrast to employing a prespecified queen or rook matrix, respectively. Similarly, our approach also leads to an RMSE reduction of around 73\% and 65\% compared to the ML estimation. Figure \ref{fig:pred} depicts $\mathrm{NO_2}$ observations and predictions obtained from all model specifications for the abovementioned central region in Mexico and for two further regions that are included for exemplary illustration. The better model performance is particularly evident in the detection of clusters of high $\mathrm{NO_2}$ concentrations, whose shapes are better captured if the spatial weights are estimated.

\section{Conclusion}\label{sec:conc}

In contrast to the classical spatial econometric models that require prior specifications of a spatial weights matrix, this paper investigates the estimation of the spatial dependence structure for spatial lattice data. A two-step adaptive lasso approach is employed to estimate the individual spatial links, assuming the sparsity of the spatial weights and the exchangeability of the random process. Moreover, we propose using cross-sectional resampling to recover the spatial dependence structure.

The Monte Carlo results demonstrate that the estimators are consistent with an increasing number of cross-sectional resampling replications. More precisely, the average percentage of correctly identified spatial connections increases, and the spatial weight estimates are most accurate for the maximum number of replications. Regarding the selection of the set of nearest locations, contradictory influences exist on the model performance. On one hand, a low number of nearest locations carries the risk of not covering all relevant neighbors. Moreover, spatial spillover effects are more concentrated on very few false neighbors, and the spatial weights estimates are more biased. On the other hand, true neighbors are better identified. In addition, more observations are available for cross-sectional resampling, and the computational effort is less demanding. Sparse dependence structures with very few true neighbors are better identified than spatial weights matrices that are more connected.

For the empirical application, we investigate the spatial dependence structure of the $\mathrm{NO_2}$ data. We find that the prediction accuracy in terms of the RMSE and the detection of clusters can be considerably improved by estimating the spatial weights in contrast to employing deterministic specifications of the $q$ nearest neighbors. This applies both to the two-step adaptive lasso approach and the classical ML estimation. However, the meaningful interpretation of standard errors with respect to the statistical significance remains an unresolved issue that should be investigated in more detail in future studies.

Finally, our approach is suitable for representing the spatial dependence structure for regular lattice data provided that the spatial dependence is constant across space. Hence, structural breaks or heterogeneous spatial dependencies are not accounted for and should be investigated in future research. Moreover, additional assumptions on the ordering of the nearest neighbors are required if other types of spatial data, such as irregular lattice or geostatistical processes, are of interest.

\end{document}